\documentclass[
twocolumn,
english,
aps,
pra,
longbibliography,
superscriptaddress,
amsmath,
amssymb,
floatfix,
noeprint,
]{revtex4-1}
\usepackage[utf8]{inputenc}
\usepackage{graphicx}
\usepackage{amsmath}
\usepackage{mathtools}
\usepackage[colorlinks=true,urlcolor=blue,citecolor=blue,linkcolor=blue]{hyperref}
\usepackage{natbib}
\usepackage{physics}
\usepackage{tikz}
\usepackage{xspace}
\usepackage{subfigure}

\usetikzlibrary{quantikz}
\usetikzlibrary{decorations.markings}

\newcommand{\eqn}[1]{Eq.~(#1)}

\newcommand{\tab}[1]{Table~#1}
\newcommand{\qedpec}{PEC+QED\xspace}
\newcommand{\nd}{\scriptscriptstyle \textrm{no-det}}   
\newcommand{\detec}{\scriptscriptstyle \textrm{det}}

\begin{document}

\title{Optimizing Symmetry Informed Probabilistic Error Cancellation}
\author{Tom O'Leary}
\affiliation{Clarendon Laboratory, University of Oxford, Parks Road, Oxford OX1
3PU, United Kingdom}
\email{thomas.oleary.research@gmail.com}
\author{Daniel J. Egger}
\affiliation{IBM Research, S\"{a}umerstrasse 4, 8803 R\"{u}schlikon, Switzerland}
\author{Dieter Jaksch}
\affiliation{Clarendon Laboratory, University of Oxford, Parks Road, Oxford OX1
3PU, United Kingdom}
\affiliation{Institute for Quantum Physics, University of Hamburg, Luruper Chaussee 149, 22761 Hamburg, Germany}

\begin{abstract}
    We show that combining quantum error detection (QED) with probabilistic error cancellation (PEC) gives more accurate and lower-variance estimates than PEC alone, provided that the symmetry measurements required for QED are carefully chosen. 
    Because noisy symmetry measurements can negate the benefits of the \qedpec approach, we cast the selection of measurement configurations as a classical optimization problem that systematically suppresses the impact of noise. 
    Applying optimized \qedpec to GHZ-state output distributions and to simulating the time-dynamics of a generalized superfast encoded Fermi-Hubbard model, we find consistent improvements over PEC. 
    For GHZ states, the optimization over symmetry measurement configurations is essential for achieving an advantage. For the Fermi-Hubbard model, \qedpec improves observable estimation on a $2 \times 2$ lattice and for larger systems the mitigation overheads can be reduced by measuring only subsets of stabilizers. 
    Our results demonstrate the importance of circuit-specific tailoring of QEM techniques and that fault-tolerant design principles may already provide value for near-term devices. 
\end{abstract}

\maketitle

\section{Introduction} \label{sec:intro}

Fault-tolerant quantum computation (FTQC) uses quantum error correction (QEC) to exponentially suppress errors arising from noisy hardware~\cite{gottesmanIntroductionQuantumError2009}.
There have been several recent experimental demonstrations of progress towards FTQC~\cite{andersenRepeatedQuantumError2020, krinnerRealizingRepeatedQuantum2022a, acharyaQuantumErrorCorrection2025}, and a steady stream of theoretical developments in reducing the associated costs~\cite{bravyiHighthresholdLowoverheadFaulttolerant2024, gidneyMagicStateCultivation2024, nguyenQuantumFaultTolerance2025}.
However, the qubit number and gate fidelity requirements for FTQC mean devices capable of fault-tolerantly executing algorithms for problems of practical interest are still likely to be several years away. 
Additionally, early fault tolerant devices will have residual error as a result of not having enough qubits or low enough physical error probabilities to make logical errors arbitrarily small. 

Quantum error mitigation (QEM) reduces errors in the output of a quantum computation without the implementation costs associated with QEC~\cite{caiQuantumErrorMitigation2023}.
QEM techniques typically require additional samples (compared to unmitigated computation), knowledge of device noise, 
and classical processing.
Numerous QEM approaches have been proposed, generally targeting observable estimation with recent extensions to sampling tasks~\cite{liuQuantumErrorMitigation2025, dutkiewiczErrorMitigationCircuit2025}. 
A subset of QEM techniques provide unbiased estimates of a target observable or distribution, most prominently probabilistic error cancellation (PEC)~\cite{temmeErrorMitigationShortdepth2017, endoHybridQuantumClassicalAlgorithms2021, bergProbabilisticErrorCancellation2022}.
As quantum circuits grow beyond the point of classical verifiability, such accuracy guarantees enable confidence in solution quality and are therefore important for scalable noise mitigation.  
Unlike QEC, QEM has no sharp error threshold before becoming useful and has been successfully applied across a broad range of experiments~\cite{kohExperimentalRealizationMeasurementInduced2023, obrienPurificationbasedQuantumError2023, kimEvidenceUtilityQuantum2023, russoTestingPlatformindependentQuantum2023, yoshiokaDiagonalizationLargeManybody2025, haghshenasDigitalQuantumMagnetism2025, fischerDynamicalSimulationsManybody2026}.
Despite these successes, QEM is limited by the exponential asymptotic scaling of sampling costs with system size and error probability~\cite{takagiFundamentalLimitsQuantum2022, quekExponentiallyTighterBounds2022, tsubouchiUniversalCostBound2023}.
This means that these techniques cannot entirely replace FTQC using QEC, which conventionally has an overhead growing poly-logarithmically in system size and time~\cite{knillResilientQuantumComputation1998} and which recent work has reduced to be constant in system size and growing near-logarithmically in time~\cite{nguyenQuantumFaultTolerance2025}.  

The question of whether QEM can bridge the gap between hardware with and without full QEC is therefore important. An affirmative answer may result in a practical quantum advantage sooner rather than later. 
Furthermore, in the fault-tolerant regime, QEM may help reduce residual logical error~\cite{suzukiQuantumErrorMitigation2022a, piveteauErrorMitigationUniversal2021,lostaglioErrorMitigationQuantumassisted2021, dutkiewiczErrorMitigationCircuit2025, tsubouchiSymmetricCliffordTwirling2025, kumarCoDesigningErrorMitigation2026}.

\begin{figure*}[ht]
  \centering
    \includegraphics[width=0.99\textwidth]{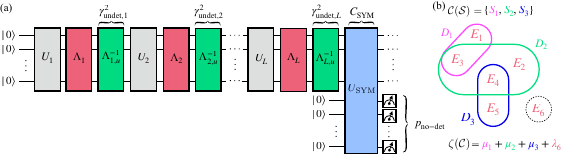}
  \caption{Overview of \qedpec. 
  (a) In a circuit with $L$-layers of noisy gates, detectable errors are mitigated by measuring symmetries using circuit $U_{\text{SYM}}$ and post-selecting, retaining circuits with probability $p_{\scriptscriptstyle \nd}$. Remaining errors are mitigated by applying PEC to cancel undetectable noise $\Lambda_{l, u}^{-1}$ at each layer $l$, incurring a sampling cost $\gamma^2_{{\scriptscriptstyle \text{undet}, l}}$ which multiplies per layer. Mitigating noise in $U_{\text{SYM}}$ introduces a sampling cost $C_{\scriptscriptstyle\text{SYM}}$.
  (b) Given a set of possible symmetries to measure $\mathcal{S} = \{S_i \}_{i=1}^4$, we seek a configuration $\mathcal{C} \subset \mathcal{S}$ with minimal cost. Each symmetry $S_i \in \mathcal{C}$ has a detectable error set $D_i$ and incurs cost $\mu_i$, while each undetected error $E_j$ incurs cost $\lambda_j$.}
  \label{fig:qed_pec_diag}
\end{figure*}

Symmetries in circuits or quantum states may help bridge this gap.
Symmetry forms the foundation of QEC using the stabilizer formalism~\cite{gottesmanStabilizerCodesQuantum1997} and is often naturally present in problems of interest.
One subsequent QEM modality targets logical noise remaining 
after imperfect universal resource state production or syndrome decoding~\cite{suzukiQuantumErrorMitigation2022a, piveteauErrorMitigationUniversal2021, lostaglioErrorMitigationQuantumassisted2021, dutkiewiczErrorMitigationCircuit2025, wahlZeroNoiseExtrapolation2023, aharonovSyndromeAwareMitigation2025, kumarCoDesigningErrorMitigation2026}.
Another modality implements QEM at the physical level, using symmetry inherent to the problem of interest~\cite{bonet-monroigLowcostErrorMitigation2018, mcardleErrorMitigatedDigitalQuantum2019, mccleanDecodingQuantumErrors2020, caiQuantumErrorMitigation2021, caiMultiexponentialErrorExtrapolation2021, kakkarCharacterizingErrorMitigation2022a, zhaoGrouptheoreticErrorMitigation2024, nigmatullinExperimentalDemonstrationBreakEven2025, liaoAchievingComputationalGains2025} 
or engineering additional symmetry~\cite{bonet-monroigLowcostErrorMitigation2018, papicNearTermFermionicSimulation2025}
to reduce QEM sampling overheads or bias.
Reducing the sampling costs of PEC with symmetry information has been studied at the logical~\cite{suzukiQuantumErrorMitigation2022a, piveteauErrorMitigationUniversal2021, lostaglioErrorMitigationQuantumassisted2021, dutkiewiczErrorMitigationCircuit2025} and physical levels~\cite{caiMultiexponentialErrorExtrapolation2021, papicNearTermFermionicSimulation2025}, where PEC mitigates the errors remaining after quantum error detection (QED). 
We refer to this approach as \qedpec.

\qedpec combines the lower sampling cost of QED with the accuracy guarantees of PEC. 
Although QEM is asymptotically limited by an exponential growth in sample complexity, the same growth strongly rewards methods which refine QEM using problem specific information.
\qedpec was initially proposed in~\cite{caiMultiexponentialErrorExtrapolation2021} and numerically tested on time evolution of the Fermi-Hubbard model with a Jordan-Wigner fermion-to-qubit mapping.
The technique has been since developed further and numerically tested across a range of fermion-to-qubit mappings, remaining effective even after accounting for additional noise from symmetry measurements~\cite{papicNearTermFermionicSimulation2025}.
However, this additional noise still limits the effectiveness of \qedpec and in other applications may preclude an advantage over PEC entirely.

Here, we optimize \qedpec by exploiting the freedom in which symmetries are used for error detection. We balance the trade-off between increased QEM cost reductions from additional symmetry measurements and decreased accuracy from errors during these measurements, in order to minimize the overall mitigation overhead.
We focus on application at the physical level, with noise modelled by a Pauli-Lindblad model, for which PEC has been experimentally validated~\cite{bergProbabilisticErrorCancellation2022, chenDisambiguatingPauliNoise2026}. 
We formulate the problem of choosing symmetry measurement configurations as a variant of the set covering optimization problem, which can admit approximation algorithms with performance guarantees. 
We find that using only a subset of the problem symmetries in \qedpec can be advantageous. 
We then numerically test our approach by applying \qedpec to GHZ state output distributions. 
Here, a careful selection of symmetry measurements is necessary for an advantage over PEC in probability distribution mitigation with growing system size. 
We also simulate the time dynamics of the spinless Fermi-Hubbard model on a square lattice. 
We find \qedpec systematically achieves a lower error than QED or PEC alone in mitigating lattice operator observables and a two-point charge correlator.  

In Sec.~\ref{sec:background} we review the background of PEC and QED.
In Sec.~\ref{sec:qedpec} we analyze the sampling costs of \qedpec and introduce our symmetry-measurement optimization framework.
Numerical investigations of GHZ state preparation are presented in Sec.~\ref{sec:ghz} and the Fermi-Hubbard model in Sec.~\ref{sec:fh}.
We discuss our results and conclude in Sec.~\ref{sec:disc}.

\section{Background} \label{sec:background}

\subsubsection{Probabilistic Error Cancellation} 
PEC views a noisy quantum gate as an ideal operation followed by a probabilistically applied error. 
With knowledge of the error probability distribution, noise is cancelled on average by sampling recovery operations from the inverse distribution and appending these operations after each noisy gate. 

We now introduce the Sparse-Pauli-Lindblad (SPL) noise model~\cite{bergProbabilisticErrorCancellation2022}, a broader introduction to PEC is in App.~\ref{app:pec}. 
The model can represent Markovian noise reflecting device locality and, in principle, can generate higher-weight correlated effects.     
For an $n$-qubit circuit which can be partitioned into $L$-layers of gates acting in parallel, this model represents noise acting after the $l$th-layer as
\begin{align} \label{eq:spl}
    \Lambda_l (\rho) = \bigcirc_{k \in \mathcal{K}_l} \left( w_{l, k} \cdot + (1 - w_{l, k}) E_{l, k} \cdot E_{l, k} \right) (\rho),
\end{align}
where the index set $\mathcal{K}_l$ is over local Pauli errors $E_{l, k}$ occurring with probability $1 - w_{l, k}$ and $\bigcirc$ denotes quantum channel composition.
The set $\{E_{l, k}\}_{k \in \mathcal{K}_l}$ are the \emph{generators} of the model and are restricted to weight-one and -two Pauli terms, which are chosen to reflect the underlying physical qubit connectivity.
Each sub-channel $w_{l, k} \rho + (1 - w_{l, k}) E_{l, k} \rho E_{l, k}$ acts on $\rho$ sequentially and each sub-channel commutes.
The probability of each generator is calculated by estimating the model \emph{weights} using a benchmarking procedure \cite{bergProbabilisticErrorCancellation2022, bergTechniquesLearningSparse2024, malekakhlaghEfficientLindbladSynthesis2025}.
The weights $\{ \lambda_{l, k} \}_{k \in \mathcal{K}_l}$ are related to the error probabilities as $w_{l, k} = (1 + \exp(-2 \lambda_{l, k})) / 2$.

The inverse SPL model of Eq.~(\ref{eq:spl})
\begin{align} \label{eq:inv_spl}
    \Lambda_l^{-1} (\rho) &= \bigcirc_{k \in \mathcal{K}_l} \frac{1}{2 w_{l, k} - 1} \left( w_{l, k} \cdot - (1 - w_{l, k}) E_{l, k} \cdot E_{l, k} \right) (\rho)\\
     &= \gamma_l \bigcirc_{k \in \mathcal{K}_l} \left( w_{l, k} \cdot - (1 - w_{l, k}) E_{l, k} \cdot E_{l, k} \right) (\rho),\\ \label{eq:gamma_l}
     \gamma_l &= \prod_{k \in \mathcal{K}_l} \frac{1}{2 w_{l, k} - 1},
\end{align}
defines a quasi-probability distribution that is sampled from to cancel the effect of noise on average.
For example, for $L=1$ we may wish to estimate the expectation value $\langle O_{\text{ideal}} \rangle = \text{tr}\{O \mathcal{U}_1(\rho) \}$, for a Pauli observable $O = \bigotimes_{i=1}^n P_i$, unitary channel $\mathcal{U}_1(\rho) = U_1\rho U_1^{\dagger}$ and state $\rho$, but $\mathcal{U}_1$ is affected by noise, giving $ \langle O_{\text{noisy}} \rangle = \text{tr}\{O \Lambda_1 \circ \mathcal{U}_1(\rho) \}$. 
We estimate $\text{tr}\{O \Lambda_1^{-1} \circ \Lambda_1 \circ \mathcal{U}_1(\rho) \} = \langle O_{\text{ideal}} \rangle$ using $M$ samples. 
For each $1 \leq j \leq M$ and $k \in \mathcal{K}_l$ sample 
$b_{jk} \in \{0,1\}$ from a Bernoulli distribution with probability $p_{l, k} = 1 - w_{l, k}$.
Define a Pauli string $R_j = \prod_{k \in \mathcal{K}_l} E_{l, k}^{b_{jk}}$ and sign $s_j =(-1)^{\sum_{k \in \mathcal{K}_l} b_{jk}}$. 
Then execute a circuit to prepare $R_j(\mathcal{U}_1(\rho))R_j$ and measure in the eigenbasis of $O$, obtaining a single-shot observable estimate $o_j$.
The set of $M$ observable estimates obtained from repeatedly applying this procedure are re-weighted by $\gamma_1 s_j$ to construct an estimator of the sample mean
\begin{align} \label{eq:pec_est}
 \overline{O}_{\scriptscriptstyle \text{PEC}} &= \frac{\gamma_1}{M} \sum_{j=1}^M s_j o_j,\\
 \mathbb{E}[ \overline{O}_{\scriptscriptstyle \text{PEC}}] &= \langle O_{\text{ideal}} \rangle.
\end{align}
The variance of $\text{Var}[\overline{O}_{\scriptscriptstyle \text{PEC}}]$ is $\gamma_1^2$ times larger than an unmitigated estimator. Therefore, $\gamma_1^2$ times more samples are needed to estimate $\mathbb{E}[O]_{\text{PEC}}$ within a fixed level of precision. 
We assume noise introduced by $R_j$ is small in comparison to $\Lambda_l$ or $R_j$ can be noiselessly implemented using frame changes \cite{mckayEfficientGatesQuantum2017}. 

Typically PEC is used when estimating an expected value
\begin{align}
    \langle O_{\text{ideal}} \rangle = \text{tr}\{O \mathcal{U}_L \circ \cdots \circ \mathcal{U}_2 \circ \mathcal{U}_1 (\rho)\},
\end{align}
for a circuit with multiple noisy layers of gates $\mathcal{U}_i$.
However, we can only estimate the noisy expected value
\begin{align}
    \langle O_{\text{noisy}} \rangle &= \text{tr}\{O \Lambda_L \circ \mathcal{U}_L \circ \cdots \circ \mathcal{U}_2 \circ \Lambda_1 \circ \mathcal{U}_1 (\rho)\}.
\end{align}
With multiple error locations $1 \leq l \leq L$, for each circuit execution 
Paulis are sampled independently for each $l$, building the set $\left\{ \{ R_{l,j} \}_{l=1}^L \right\}_{j=1}^M$. The total sign and mitigation overheads are then $s^{\scriptscriptstyle\text{tot}}_j = \prod_{l=1}^M s_l$ and $\gamma^{\scriptscriptstyle\text{tot}} = \prod_{l=1}^L\gamma_l$. 
The overall mitigation cost for $L$-layers is
\begin{align}
    C_{\scriptscriptstyle \text{PEC}} &= \prod_{l=1}^L\gamma_l^2\\
    &= e^{4 \sum_{l=1}^L \sum_{k \in \mathcal{K}_l} \lambda_{l, k}}.
\end{align}\label{eq:pec_cost}
Being able to exactly calculate $C_{\scriptscriptstyle \text{PEC}}$ as a simple sum of model weights is an advantage enabled by considering Pauli-Lindblad noise models, which we make use of when minimizing $C_{\scriptscriptstyle \text{PEC}}$ in Sec.~\ref{sec:sym_selec}.

\subsubsection{\label{sec:qed} Quantum Error Detection}
QED discards executions with detected errors to improve accuracy. 
In principle, QED can be applied to any circuit by embedding it in an error-detecting code~\cite{nielsenQuantumComputationQuantum2010}.
In the context of QEM, this can be viewed as symmetry-based post-selection: a circuit run is discarded if there is a discrepancy between a measured symmetry and its expected value.
QEM methods that enforce symmetry constraints inherent to a problem in this way are often referred to as symmetry verification.
Several such approaches check Pauli symmetries and apply post-selection \cite{bonet-monroigLowcostErrorMitigation2018, mcardleErrorMitigatedDigitalQuantum2019}.
To connect these ideas to a stabilizer-based error-detection framework, we focus on states with Pauli symmetries.

A quantum state $\ket{\psi}$ is stabilized by a set of Pauli operators $\mathcal{S} = \{ S_i \}_{i=1}^m$ if
\begin{align}
    S_i \ket{\psi} = s_i \ket{\psi},
\end{align}
with $s_i = \pm1$.
Here each $S_i$ is a symmetry operator whose measurement on the ideal state returns a fixed eigenvalue $s_i$.
Regions of a quantum circuit that preserve these stabilizer relations can be used to detect errors via changes in the measured eigenvalues.
Errors that commute with all stabilizers leave all symmetry measurement outcomes unchanged and are therefore undetectable.
Correspondingly, detectable errors anti-commute with at least one stabilizer element.
This anti-commutation relation can be checked by measuring $s_i = \langle S_i \rangle$ because
\begin{align}
    \bra{\psi}E S_i E \ket{\psi} = (-1)^{\langle S_i, E\rangle_{\text{sp}}}s_i,
\end{align}
where $\langle P_i, P_j\rangle_{\text{sp}}$ is the symplectic inner product, which equals 0 if $[P_i, P_j] = 0$ and 1 otherwise.
Thus, if an error anti-commutes with $S_i$, it flips the corresponding measurement outcome.
Comparing the measured values $\{s_i\}_{i=1}^m$ to either a fixed target set or to values obtained in a previous measurement round enables inference of whether an error has occurred.
By running a circuit multiple times and discarding executions where errors are detected, a smaller set of more accurate circuit outputs can be obtained, which is the essence of QED as an error-mitigation strategy.

The QED estimator for an observable $O$ is then
\begin{align} \label{eq:qed_est}
 \overline{O}_{\scriptscriptstyle\text{QED}} &= \frac{1}{M_{\scriptscriptstyle \text{no-det}}} \sum_{j=1}^{M_{\scriptscriptstyle \text{no-det}}} o_j,
\end{align}
where the sum runs over the $M_{\scriptscriptstyle \text{no-det}}$ shots in which errors were not detected.
Estimators obtained using QED will generally be biased by undetectable errors, which includes pairs of detectable errors which anti-commute with the same subset of $\mathcal{S}$, and by state-preparation-and-measurement errors.
The sampling cost of QED with post-selection is
\begin{align}
C_{\scriptscriptstyle \text{QED}} = \frac{1}{p_{\scriptscriptstyle \nd}},
\end{align}
where $p_{\scriptscriptstyle \nd}$ is the probability that no error is detected: either no error occurs or an undetectable error occurs.
Example calculations of $p_{\scriptscriptstyle \nd}$ are in App.~\ref{app:qed}.

\section{Reducing PEC sampling costs with symmetry-information} \label{sec:qedpec}
The sampling overhead of PEC results from only having a statistical view of hardware errors instead of knowing precisely which errors occur during runtime.
This suggests that we could lower the sampling cost by reducing the uncertainty about which errors occur during circuit execution.

\subsection{Implementation and Sampling Cost}
We now describe how ignoring errors removed by post-selection in QED reduces the sampling overhead of PEC.
We first mitigate a single layer of noisy gates. We then consider circuits with symmetry preserving layers and circuits where symmetry preserving layers are decomposed into 
layers of hardware native basis gates, which may not individually preserve symmetry. 
An overview of \qedpec is presented in Fig.~{\ref{fig:qed_pec_diag}}(a).

For an SPL noise model as in Eq.~(\ref{eq:spl}) and a set of symmetries $\mathcal{S}$, we define the set of undetectable error generators at the $l$th-layer as 
\begin{align}
   \mathcal{K}^{\scriptscriptstyle \text{undet}}_l = \{k \in \mathcal{K}_l \ \lvert \ [E_{l, k}, S_i]=0 \ \forall \ S_i \in \mathcal{S}, \ E_{l, k} \notin \mathcal{S} \},
\end{align}
and the set of detectable error generators as 
\begin{align}
   \mathcal{K}^{\scriptscriptstyle \text{det}}_l = \{k \in \mathcal{K}_l \ \lvert \ k \notin  \mathcal{K}^{\scriptscriptstyle \text{undet}}_l \}.
\end{align}
By discarding circuit executions where an error is detected and mitigating the remaining circuits,
we apply PEC to errors 
\begin{align}
    \Lambda^{\scriptscriptstyle \text{undet}}_l (\rho) = \bigcirc_{k \in \mathcal{K}_l^{\scriptscriptstyle \text{undet}}} \left( w_{l, k} \cdot + (1 - w_{l, k}) E_{l, k} \cdot E_{l, k} \right) (\rho),
\end{align}
generated by $\mathcal{K}^{\scriptscriptstyle \text{undet}}_l$ only.
This model does not include contributions from pairs of detectable errors and is therefore a 1st-order approximation.
Using Eq.~(\ref{eq:pec_est}) and Eq.~(\ref{eq:qed_est}) we can then write the sample estimator for \qedpec as 
\begin{align}
    \overline{O}_{\scriptscriptstyle \textrm{\qedpec}} &= \frac{\gamma^{\scriptscriptstyle \text{undet}}_l}{M_{\nd}} \sum_{j=1}^{M_{\nd}} s_j o_j,
\end{align}
where $\gamma^{\scriptscriptstyle \text{undet}}_l = \prod_{k \in \mathcal{K}^{\scriptscriptstyle \text{undet}}_l} (2 w_{l, k} - 1)^{-1}$.
The estimator mean squared error (MSE) is
\begin{align}\label{eq:qedpec_mse}\nonumber
    \textrm{MSE}[\overline{O}_{\scriptscriptstyle \textrm{\qedpec}}] &= \textrm{Bias}[\overline{O}_{\scriptscriptstyle \textrm{\qedpec}}]^2 + \\  &\textrm{Var}[\overline{O}_{\scriptscriptstyle \textrm{\qedpec}}],
\end{align}
where
$\text{Bias}[\overline{O}_{\scriptscriptstyle \textrm{\qedpec}}] = \mathbb{E}[\overline{O}_{\scriptscriptstyle \textrm{\qedpec}}] - \langle O_{\text{ideal}} \rangle$ and $\text{Var}[\overline{O}_{\scriptscriptstyle \textrm{\qedpec}}] =  C_{\scriptscriptstyle \textrm{\qedpec}} \text{Var}[\overline{O}]$.
The bias will have contributions from i) errors introduced during QED measurement ii) errors introduced by PEC recovery operations and iii) higher order undetectable errors.  
For a single layer of gates the sampling overhead is
\begin{align}
    C_{{\scriptscriptstyle \text{PEC+QED}}, l} = \frac{(\gamma^{\scriptscriptstyle\text{undet}}_l)^2}{p_{{\nd}, l}},
\end{align}
where $p_{{\nd}, l}$ is the probability that no error is detected after the layer.

The case of a circuit with a single layer of gates can be extended to a circuit with multiple symmetry preserving layers and symmetry measurements at the end
\begin{align}
    U_L \circ \cdots \circ U_2 \circ U_1 (\rho),\\
    S_i U_l(\rho) = U_l(S_i\rho) \ \forall \ (i, l).
\end{align} 
In this case, a single detectable or undetectable error will remain as such whether the relevant symmetries are measured immediately after the error occurred or at the end of the circuit \cite{caiMultiexponentialErrorExtrapolation2021}. This is because each symmetry can be freely propagated through the circuit.
Therefore, the target of PEC in \qedpec is the undetectable noise at each layer $\Lambda_{\text{undet}, l}$.
The total sampling cost with ideal symmetry measurements for an $L$-layer circuit is  
\begin{align}
    C_{\scriptscriptstyle \text{PEC+QED}} &= \frac{\left(\gamma_{\scriptscriptstyle\text{undet}}\right)^2}{p_{\nd}}, 
\end{align}
where $\gamma_{\scriptscriptstyle \text{undet}} = \prod_{l=1}^L \gamma^{ \scriptscriptstyle\text{undet}}_l$ and the absence of a layer index $l$ denotes a cost over the whole circuit.

We now describe how \qedpec applies to general circuits.
In practice, circuits are decomposed into a set of noisy hardware-compatible basis gates.
While the whole circuit, or circuit sub-regions, may commute with the symmetry operator, the noisy layers mitigated by PEC may not. 
An error which is undetectable at the point of occurrence may propagate to become detectable. 
Therefore to correctly construct $\Lambda^{\scriptscriptstyle \text{undet}}_l$, it is necessary to propagate each error in $\Lambda_{l}$ through the circuit to the point of symmetry measurement. 

Errors are efficiently propagated through Clifford layers using a classical computer \cite{gottesmanHeisenbergRepresentationQuantum1998}.
Non-Clifford layers will convert a Pauli error into a sum of Pauli errors which in general grows exponentially fast with circuit depth.
However, propagation can be performed efficiently for certain circuits.
As shown in Ref.~\cite{papicNearTermFermionicSimulation2025}, this is the case for layers of Pauli rotation gates $R_P(\theta) = \exp(-\mathrm{i} \theta P)$ if, in the absence of noise, the input state has the correct symmetry and the circuit remains in the same symmetry sector for any $\theta$. This is because $R_P(\theta) E = E R_P((-1)^{\langle E, P \rangle}\theta)$, where $\langle E, P \rangle = \pm1$.
This propagation may flip $\theta \rightarrow -\theta$, however, because the circuit remains in the same symmetry sector for any $\theta$, the outcome of QED will only be affected by $E$. 
Therefore it suffices to propagate $E$ through Clifford operations. 
For more detail see App.~\ref{app:error_prop}.

There is some freedom in how far through a circuit errors are propagated.
An error $E$ can be propagated up to $U_{\scriptscriptstyle \textrm{SYM}} = U_{S_m} \cdots U_{S_2} U_{S_1}$, the circuits which measure $\{ s_i \}_{i=1}^m$, to become $E'$. 
Detectability can then be checked by computing $[E', S_i] \ \forall \ S_i \in \mathcal{S} $, see Sec.~\ref{sec:qed}. 
Alternatively, a more accurate description of undetectable noise is obtained by taking the structure of each symmetry measurement circuit into account.
We assume $U_{S_i}$ consists of Clifford gates and ancilla measurements $\langle Z_{\text{anc}(i)} \rangle$ for each $S_i$, then propagate $E'$ up to each ancilla measurement to give $E'_i$ and check $[ E'_i, Z_{\text{anc}(i)}]$. 
This also enables errors within $U_{S_i}$ to be classified as detectable and then mitigated using \qedpec, for more detail see App.~\ref{app:sym_circ_mit}.

\subsection{Optimizing Symmetry Measurements} \label{sec:sym_selec}
Given a set of multiple possible symmetries to measure, we would like to select those that minimize the sampling cost of \qedpec.
Each additional symmetry measurement may reduce $\gamma_{\scriptscriptstyle \text{undet}}$ at the cost of an increase in bias from undetectable errors and variance from false detections due to noise in the symmetry measurement circuits, $U_{\scriptscriptstyle \textrm{SYM}}$.
Ideally, we would choose a configuration which directly minimizes Eq.~(\ref{eq:qedpec_mse}), however, $\text{Bias}[\overline{O}_{\scriptscriptstyle \qedpec}]$ is not practical to compute as it requires knowing the noise-free observable $\langle O \rangle$.
Instead, we minimize the variance of Eq.~(\ref{eq:qedpec_mse}), using two alternative approaches for approximating the impact of noisy symmetry measurements.
First, we assume that additional QEM can be applied to $U_{\scriptscriptstyle \textrm{SYM}}$ such that the mitigated symmetry circuits introduce no estimator bias. 
Second, we focus on experiments where symmetries are measured after the observable to be mitigated, which we refer to as deferred, and don't directly contribute to bias. 

We setup the optimization by fixing a set of candidate stabilizer checks $\mathcal{S} = \{ S_i \}_{i=1}^M$ and look for a subset of checks $\mathcal{C} \subseteq \mathcal{S}$ to execute.
Circuit noise is modeled as an ensemble of Pauli errors and weights
\begin{align}
    \mathcal{N} = \bigsqcup_{l=1}^L \{ (E_{l, k'}, \lambda_{l, k'} ) \}_{k' \in \mathcal{K}_l},
\end{align}
where $l$ runs over gate layers and $k'$ runs over the generators of the noise model for each layer, each error $E_{l, k'}$ has a model weight $\lambda_{l, k'}$.
For notational simplicity, we flatten the error indices $E_{l, k'} \rightarrow E_k$ and write $( E_k, \lambda_k) \in \mathcal{N}$.
Let $E'_k$ be each error $E_k$ propagated to the symmetry-measurement point. 
The set of error indices detected by stabilizer $S_i$ is
\begin{align}
    D_i = \{ \, k : (E_k, \lambda_k) \in \mathcal{N} \,\vert \, \{E'_k, S_i \} = 0 \},
\end{align}
so an error is undetected by $\mathcal{C}$ if it commutes with all $S_i \in \mathcal{C}$.
An error $E_k$ not covered by the selected stabilizers introduces a cost of $\lambda_k$.
The set of error indices detected by $\mathcal{C}$ is
\begin{align}
    \mathcal{D} (\mathcal{C}) = \bigcup_{i: S_i\in\mathcal{C}} D_i.
\end{align}
Defining $\mathcal{K} = \bigsqcup_{l=1}^L \mathcal{K}_l$, the cost of removing undetectable errors with \qedpec is then
\begin{align} \label{eq:cost_pec_undet}
   \left( \gamma_{\scriptscriptstyle\text{undet}} \right)^2  = e^{4 \sum_{k \in \mathcal{K} \backslash \mathcal{D} (\mathcal{C})}\lambda_k },
\end{align}
which represents the benefits of measuring additional symmetries.
Our two approaches to the optimization then differ in how they formulate the penalty for additional measurements.

\subsubsection{\label{sec:sym_selec_a} Optimization with symmetry circuit mitigation}
Here we assume that symmetry measurement circuits are mitigated with PEC such that they introduce no additional bias~\cite{guptaProbabilisticErrorCancellation2024}.We then minimize the total quasi-probability cost of \qedpec. 
Let the cost to select $S_i$ be the magnitude of its noise
\begin{align}
   \mu_i = \sum_{ (E_{i, q}, \lambda_{i, q}) \in \mathcal{Q}_i} \lambda_{i, q},
\end{align}
where $\mathcal{Q}_i =  \{ (E_{i, q}, \lambda_{i, q} ) \}_q$ is the noise associated with the circuit to measure $S_i$.
The cost of mitigating $U_{\scriptscriptstyle \textrm{SYM}}$ for a given stabilizer configuration $\mathcal{C}$ is then 
\begin{align} \label{eq:pec_cost_usym}
  C_{\scriptscriptstyle \textrm{SYM}} (\mathcal{C})= e^{4\sum_{i: S_i \in \mathcal{C}} \mu_i}.  
\end{align}
The total PEC cost of \qedpec is then
\begin{align}
   \left( \gamma_{\scriptscriptstyle \text{undet}} \right)^2 C_{\scriptscriptstyle \textrm{SYM}}(\mathcal{C}) = e^{4 \left( \sum_{k \in \mathcal{K} \backslash \mathcal{D} (\mathcal{C})}\lambda_k + \sum_{i: S_i\in\mathcal{C}}\mu_i \right)}.
\end{align}
Because $e^x$ is monotonic, we can define the minimization as
\begin{align} \label{eq:set_cov_1}
    & \quad \quad \underset{\mathcal{C} \subseteq \mathcal{S}}{\arg\min} \;\zeta(\mathcal{C}),\\
    \zeta(\mathcal{C}) &= \sum_{k \in \mathcal{K} \backslash \mathcal{D} (\mathcal{C})} \lambda_k + \sum_{i: S_i \in \mathcal{C}} \mu_i.
\end{align}
This is a constrained set covering problem.
Here, one must choose a collection of sets so that every element appears in at least one set, while paying a cost for each set and a penalty for each element left out, so as to minimize the set costs and penalties.
A high-level overview of the optimization can be found in Fig.~{\ref{fig:qed_pec_diag}}(b).
Finding the resulting set cover is NP-hard~\cite{lund1994hardness}.
Exact solutions can be found for moderate problem sizes using mixed-integer linear programming and standard approximation algorithms can be applied for large instances.
Here, greedy algorithms can achieve a near-optimal approximation ratio of $1 + \ln(\max_i \lvert D_i \rvert )$, since the benefit term $\sum_{k \in \mathcal{K} \backslash \mathcal{D} (\mathcal{C})} \lambda_k$ is submodular~\footnote{A submodular function $f$ has that, for $A \subseteq B \subseteq \mathcal{C}$ and $s \in  \mathcal{C} \ \backslash \ B$, $f(B \cup x) - f(B) \leq f(S \cup x) - f(A)$~\cite{doi:10.1137/090779346}} i.e. adding more symmetries gives diminishing returns and the penalty $\sum_{i: S_i \in \mathcal{C}} \mu_i$ is linear~\cite{chvatal1979greedy,wolsey1982analysis, lund1994hardness, feigeThresholdLnApproximating1998, SVIRIDENKO200441, guptaALocal2023}.

In treating the penalty as linear we assume that the cost of each symmetry measurement is fixed, and does not depend on which other symmetries are measured alongside it.
This may overestimate the noise contribution from each measurement, for example when idling noise is significant, as it does not consider circuit parallelization. 
We now describe an additional optimization which does take parallelization into account, in order to discover lower cost configurations.

The extended procedure begins by solving Eq.~(\ref{eq:set_cov_1}) with idling noise excluded from each penalty term $\mu_i$, returning a configuration $\mathcal{C}^*$.
This leverages the strong performance of greedy algorithms for set-cover problems.
Next we search for the subset of $\mathcal{C}^*$ with smallest mitigation cost when idling noise is considered.
Parallelization is taken into account by calculating $C_{\scriptscriptstyle \textrm{idle}}(\mathcal{T})$, the additional cost of mitigating idling noise in the compiled measurement circuit for a subset $\mathcal{T} \subseteq \mathcal{C}^*$.
The idling-aware objective for $\mathcal{T} \subseteq \mathcal{C}^*$ is then
\begin{align} \label{eq:idle_obj}
    \zeta_{\scriptscriptstyle \textrm{idle}}(\mathcal{T}) = \sum_{k \in \mathcal{K} \backslash \mathcal{D} (\mathcal{T})} \lambda_k + \sum_{i: S_i \in \mathcal{T}} \mu_i + C_{\scriptscriptstyle \textrm{idle}}(\mathcal{T}).
\end{align}
Because $C_{\scriptscriptstyle \textrm{idle}}(\mathcal{T})$ depends on the joint circuit layout, it cannot in general be decomposed as a sum over individual stabilizers.
This breaks the submodularity and linearity structure of the benefit and penalty terms respectively, so the greedy approximation guarantee for Eq.~(\ref{eq:set_cov_1}) no longer applies to Eq.~(\ref{eq:idle_obj}).
However, the greedy solution to Eq.~(\ref{eq:set_cov_1}) typically returns a small pool $|\mathcal{C}^*| \ll |\mathcal{S}|$, since the algorithm stops once no stabilizer offers positive net gain.
Therefore, we can often evaluate Eq.~(\ref{eq:idle_obj}) exactly over all $2^{|\mathcal{C}^*|}$ subsets $\mathcal{T} \subseteq \mathcal{C}^*$ and select the configuration $\mathcal{T}^*$ with minimum $\zeta_{\scriptscriptstyle \textrm{idle}}$.
For cases where enumeration is infeasible, a greedy algorithm can still be applied directly to $\zeta_{\scriptscriptstyle \textrm{idle}}$, albeit without a formal approximation guarantee.

\subsubsection{Optimization with deferred symmetry measurements}
If symmetries are measured after the observable, noise in the symmetry circuits cannot directly bias the observable estimator. 
However, errors during measurement circuits may cause false detections
which will increase the sample bias of the \qedpec estimator. Alternatively, falsely discarded circuit executions without errors or with undetectable errors will also increase the sample variance of the \qedpec estimator. 
With enough noise introduced by symmetry measurements, the overall bias and variance increase from false detections will outweigh the decrease in PEC cost from detecting more errors. 

To quantify this we consider an alternative penalty term for the stabilizer-selection optimization.
We approximate the detectable error probability in the main circuit and measurement circuits to first order in physical error probability
\begin{align}
    p_{\scriptscriptstyle \textrm{det}} (\mathcal{C}) \approx \sum_{k \in \mathcal{D} (\mathcal{C})} 1 - w_k, 
\end{align}
where $w_k$ is defined as in Eq.~(\ref{eq:spl}) with flattened indices and $D_i$ now includes errors introduced by symmetry measurement circuits in the configuration $\mathcal{C}$.
Then the cost of post-selection to first order, including false detections is
\begin{align}
    C_{\scriptscriptstyle \textrm{QED}}(\mathcal{C}) &= \frac{1}{1 -p_{\scriptscriptstyle \textrm{det}} (\mathcal{C})},\\
    &= e^{-\ln(1 - p_{\scriptscriptstyle \textrm{det}} (\mathcal{C}))},\\
    &\approx e^{\sum_{k \in \mathcal{D} (\mathcal{C})} 1 - w_k}.
\end{align}
We then use this sampling cost estimate as a penalty for a given stabilizer configuration. 
In this case the optimization becomes
\begin{align} \label{eq:opt_fd}
    & \quad \quad \underset{\mathcal{C} \subseteq \mathcal{S}}{\arg\min} \;\zeta(\mathcal{C}),\\
    \zeta(\mathcal{C}) &= 4 \sum_{k \in \mathcal{K} \backslash \mathcal{D} (\mathcal{C})} \lambda_k + \sum_{k \in \mathcal{D} (\mathcal{C})} 1 - w_k.
\end{align}
In contrast to Eq.~(\ref{eq:set_cov_1}) this optimization now involves a difference of submodular set functions, therefore the greedy approximation guarantees no longer apply and we must rely on heuristic methods \cite{10.5555/3020336.3020387, 10.5555/3020652.3020697}.

\subsection{Encoding to enable \qedpec}
For problems without inherent symmetry, encoding using a QED code may enable a sampling advantage over unencoded PEC.
This has been previously explored from the perspective of applying PEC to logical noise channels \cite{suzukiQuantumErrorMitigation2022a, dutkiewiczErrorMitigationCircuit2025}.
Here we consider applying PEC at the physical level, motivated by empirical evidence that physical-level noise can be effectively characterized \cite{kimEvidenceUtilityQuantum2023}.
In this case implementing QED may require an $n'$ qubit circuit with $n' > n$ and each circuit layer $U_l$ being implemented by $L_l$ noisy sub-layers acting on the $n'$ qubit space
\begin{align}
    \overline{U}_l = \Lambda_{l, L_l} \circ V_{l, L_l} \circ \cdots \circ \Lambda_{l, 2} \circ V_{l, 2} \circ \Lambda_{l, 1} \circ V_{l, 1},
\end{align}
where the overline refers to the encoded circuit.  

After $L$ layers QED is performed by measuring a set of circuit symmetries. 
Noise after each sub-layer is generated by $\overline{\mathcal{K}}_{l, d}$ and can be grouped into undetectable and detectable sets. 
Applying PEC to the undetectable group $\Lambda^{\scriptscriptstyle\textrm{undet}}_{l, d}$ contributes $(\gamma^{\scriptscriptstyle\textrm{undet}}_{l, d})^2$ to the total sampling cost for the encoded circuit
\begin{align}
    \overline{C}_{\scriptscriptstyle \text{PEC+QED}} &= \frac{\prod_{l=1}^L \prod_{d = 1}^{L_l} (\gamma^{\scriptscriptstyle\textrm{undet}}_{l, d})^2}{\overline{p}_{\nd}} \\
    &= \frac{e^{4\sum_{l=1}^L \sum_{d = 1}^{L_l} \sum_{k \in \overline{\mathcal{K}}^{\scriptscriptstyle \text{undet}}_{l,d}}\lambda_{l, d, k}}}{\overline{p}_{\nd}},
\end{align}
where $\overline{p}_{\nd}$ is the probability of not detecting an error after the encoded circuit.
Simplifying by 
assuming constant error model weights $\lambda_{l, d, k} = \lambda$, constant sub-layer depth $L_l = L'$ and  fixed size $\lvert \overline{\mathcal{K}}^{\scriptscriptstyle\text{undet}}_{l, d} \rvert = \lvert \overline{\mathcal{K}}^{\scriptscriptstyle\text{undet}}\rvert$ and $\lvert \mathcal{K}_{l} \rvert = \lvert \mathcal{K} \rvert$, for PEC on the unencoded system, we compare mitigation costs
\begin{align}
    \overline{\eta}_{\scriptscriptstyle \text{uniform}} &= \frac{C_{\scriptscriptstyle \text{PEC}}}{\overline{C}_{\scriptscriptstyle \text{PEC+QED}}^{\scriptstyle \textrm{enc}}} \\ &=  \label{eq:enc_ratio} \frac{e^{4 L \lambda (\lvert \mathcal{K} \rvert -  L' \lvert \overline{\mathcal{K}}^{\scriptscriptstyle\text{undet}}\rvert)} \overline{p}_{\nd}}{C_{\scriptscriptstyle\textrm{SYM}}}.
\end{align}
As $\overline{\eta}_{\scriptscriptstyle \text{uniform}} > 1$ is needed for a sampling cost advantage using \qedpec, $\lvert \overline{\mathcal{K}}^{\scriptscriptstyle\text{undet}}\rvert < \lvert \mathcal{K} \rvert / L'$ defines a necessary condition on the magnitude of undetectable errors in the encoded system and the increased depth needed to implement encoded gates. 
Often the generators in $\overline{\mathcal{K}}$ are weight-1 or -2 Paulis. 
If in this simplified setting a detectable error does not propagate to become undetectable, the distance of the code used for encoding can inform $\eta^{\scriptscriptstyle \textrm{enc}}_{\scriptscriptstyle \text{uniform}}$.
For code distances $\geq 5$ we can detect any single or two qubit error in $\overline{\mathcal{K}}$ and therefore $\lvert \overline{\mathcal{K}}^{\scriptscriptstyle\text{undet}}\rvert = 0$.
Although this minimizes the quasi-probabilistic aspect of the \qedpec cost, additional detections will decrease $\overline{p}_{\nd}$, and achieving a high distance may increase the complexity of the symmetry-measurement circuit $U_{\scriptscriptstyle \text{SYM}}$ and the associated mitigation cost $C_{\scriptscriptstyle \text{SYM}}$.
Therefore Eq.~(\ref{eq:enc_ratio}) may be maximized for a lower distance encoding with a lower detection rate and less complex $U_{\scriptscriptstyle \text{SYM}}$.

\section{Numerical Experiments} \label{sec:numerics}
We now describe the numerical experiments performed to test the effectiveness of \qedpec in comparison to PEC and QED individually.

\subsection{GHZ state output distributions with mitigated sampling} \label{sec:ghz}
Greenberger–Horne–Zeilinger (GHZ) states for $n$-qubits are entangled quantum states
\begin{align}
    \ket{\psi} &= \frac{1}{\sqrt{2}}\left(\ket{0}^{\otimes n} + \ket{1}^{\otimes n}\right).
\end{align}
These states are broadly applicable, for example in quantum metrology~\cite{giovannettiAdvancesQuantumMetrology2011} and QEC~\cite{shorFaulttolerantQuantumComputation1997, bartolucciFusionbasedQuantumComputation2021}.
However, GHZ states are particularly sensitive to noise, increasingly so as $n$ grows~\cite{monz14QubitEntanglementCreation2011}. 
GHZ states are stabilized by elements of the group generated by
\begin{align}
    \mathbb{G}_{\scriptscriptstyle\textrm{GHZ}} &=\left\{ Z_0 Z_1, Z_1 Z_2, \dots ,Z_{n-2} Z_{n-1}, X^{\otimes n}\right\},
\end{align}
where $Z_i Z_{i + 1}$ and $X^{\otimes n}$ are $Z$- and $X$-type generators respectively. Performing QED with $\mathbb{G}_{\scriptscriptstyle\textrm{GHZ}}$ enables error-mitigated GHZ state preparation~\cite{liaoAchievingComputationalGains2025}.

\begin{figure}
    \centering
    \includegraphics[width=0.9\linewidth]{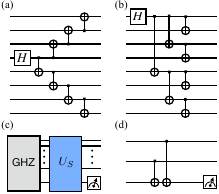}
    \caption{(a) GHZ state preparation on $n = 8$ qubits with centre-out CNOT fan from qubit 3.
    (b) GHZ state preparation on $n = 8$ qubits with logarithmic depth \cite{mooneyGenerationVerification27qubit2021}.
    (c) GHZ state preparation and stabilizer measurement circuit.
    (d) Example circuit for measuring a $ZZ$ stabilizer.
    }
    \label{fig:ghz_circs}
\end{figure}

We focus on a GHZ state preparation circuit with linear depth and one with logarithmic depth in $n$, shown for $n = 8$ qubits in Fig.~\ref{fig:ghz_circs}(a) and (b).
Preparation circuits are followed by measurements of stabilizers from $\mathbb{G}_{\scriptscriptstyle\textrm{GHZ}}$ for QED-based QEM, as shown in Fig.~\ref{fig:ghz_circs}(c) and (d).
We use a single ancilla qubit for each stabilizer measurement. 
After stabilizer measurement, all non-ancilla qubits are measured in the computational basis.

We focus on sampling from the mitigated circuit output probability distribution with increasing GHZ state size $n$. 
For PEC we use the following procedure (see Ref.~\cite{liuQuantumErrorMitigation2025} for more detail).
As with PEC for observable estimation, each circuit execution has an associated set of recovery Paulis, a sign $y_i \in \{\pm1 \}$ and a measured bitstring $x_i$.
The mitigated distribution is then constructed by accumulating signed counts: $p(x) \propto \sum_i y_i \delta(x_i, x)$.
Cancellation between the positive and negative contributions reduces the effective number of samples below the shot count, leading to an increased sampling overhead. 
QED is implemented by discarding counts which do not pass post-selection. 
\qedpec is then implemented by applying QED to the PEC-mitigated counts, where PEC is applied only to undetectable errors

We use the total square error (TSE) between the noise-free and noisy distributions to quantify accuracy, which for probability distributions $p$ and $q$ over bitstrings $z$, is defined as
\begin{align}
    \text{TSE}(p,q) &= \sum_z (p(z) - q(z))^2.
\end{align}
\textit{Noise model}. 
We add noise described by a local-Pauli error model.
After two-qubit gates we use an SPL model with total error probability $p$, with generators consisting of all weight-one and weight-two Pauli strings. 
For one-qubit operations we use an SPL model with uniform total error probability and take weight-one Pauli strings as generators. 
After one-qubit unitary gates (excluding $R_Z(\theta)$, which is assumed noise-free~\cite{mckayEfficientGatesQuantum2017}) and during idling the total error probability is $p / 10$, before measurements $p$ and after resets $2 p$.
Idling noise is applied to each inactive qubit at each circuit layer.
We assume all-to-all circuit connectivity to isolate the performance of different QEM techniques from an architecture dependent SWAP overhead. 
We provide more detail in App.~\ref{app:sim}.

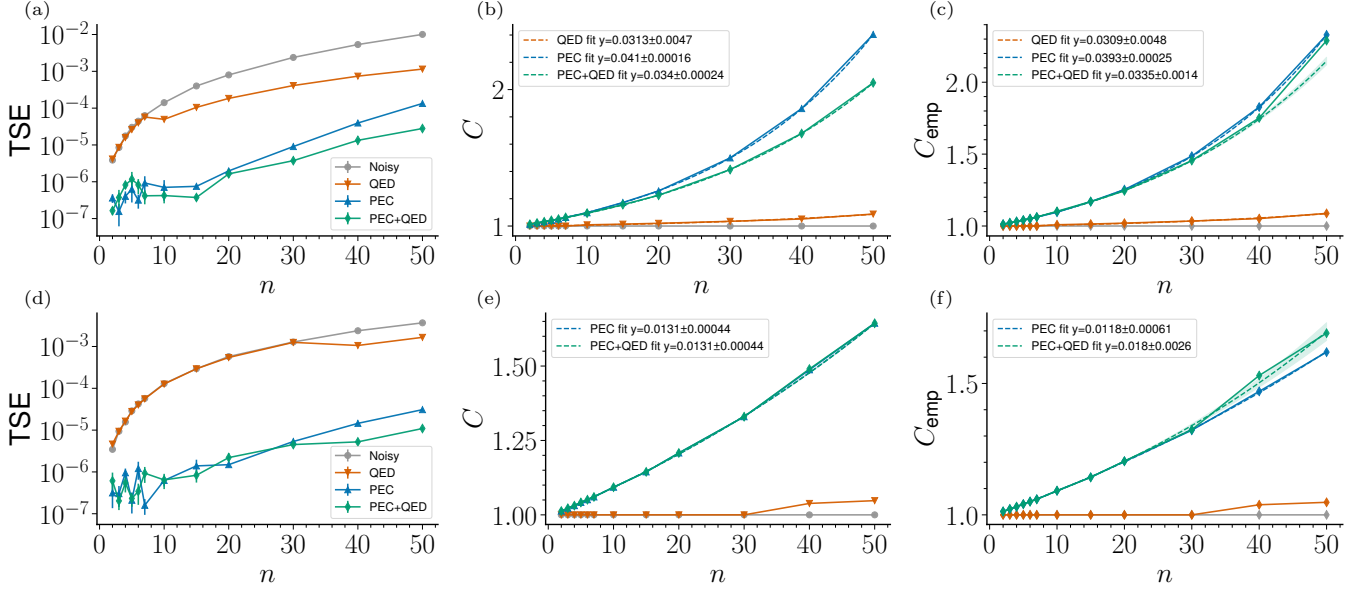
\begin{figure*}[ht]
  \centering
  \begin{tabular}{@{}ccc@{}}
    \includegraphics[width=0.32\textwidth]{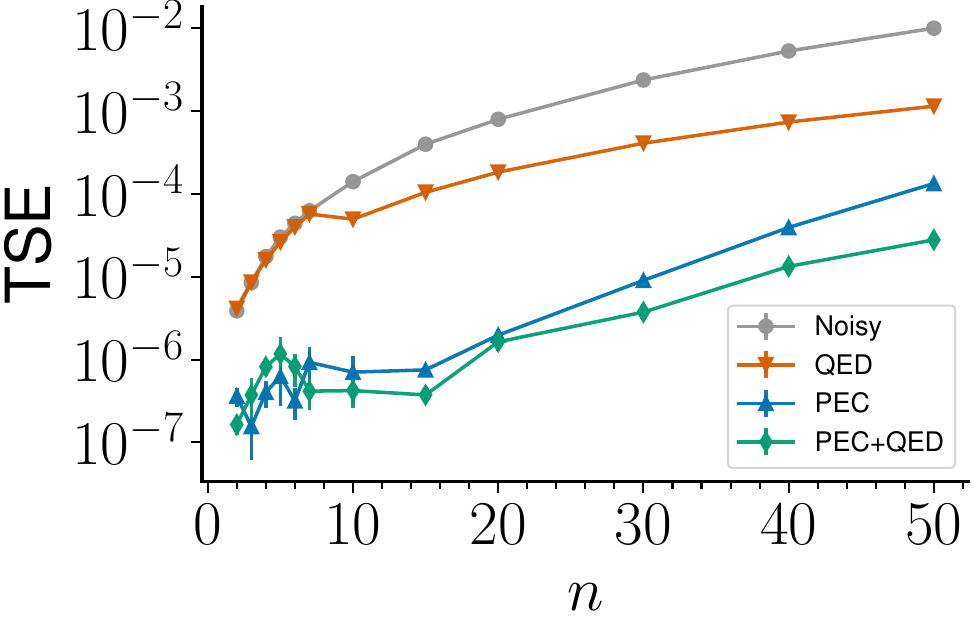} 
    \begin{picture}(0,0)
        \put(-160,108){\scriptsize (a)} 
    \end{picture}
    
    &
    \includegraphics[width=0.32\textwidth]{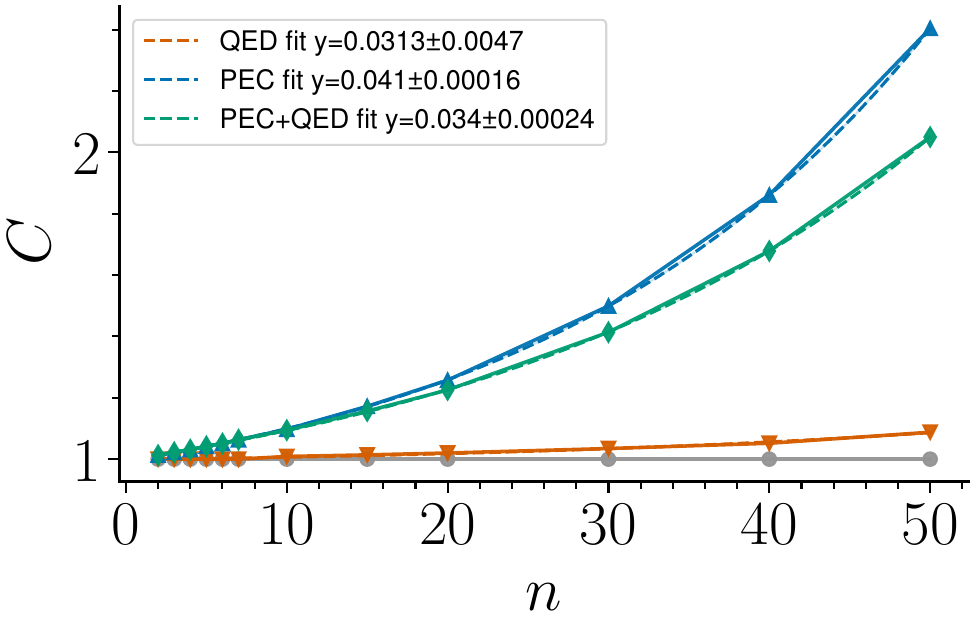} 
    \begin{picture}(0,0)
        \put(-160,108){\scriptsize (b)} 
    \end{picture}
    
    &
    \includegraphics[width=0.32\textwidth]{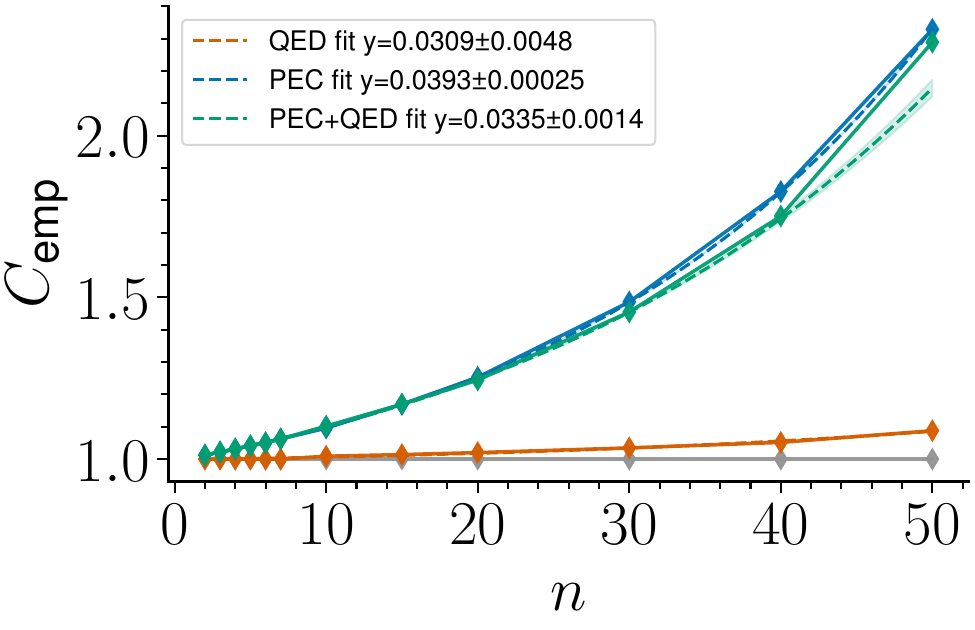}
    \begin{picture}(0,0)
        \put(-160,108){\scriptsize (c)} 
    \end{picture}
   \\
    \includegraphics[width=0.32\textwidth]{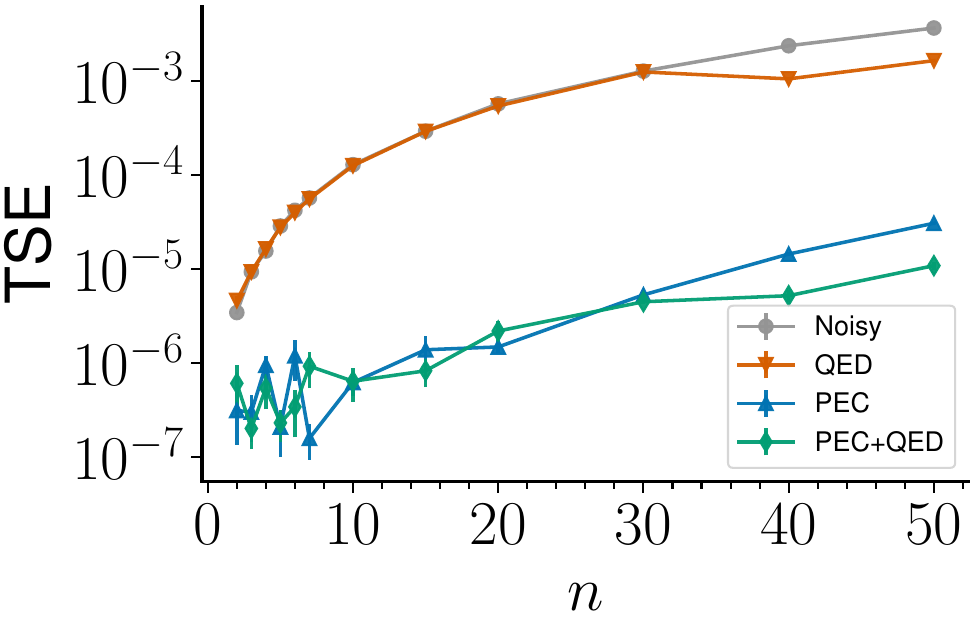}     
    \begin{picture}(0,0)
        \put(-160,108){\scriptsize (d)} 
    \end{picture}
    
    &
    \includegraphics[width=0.32\textwidth]{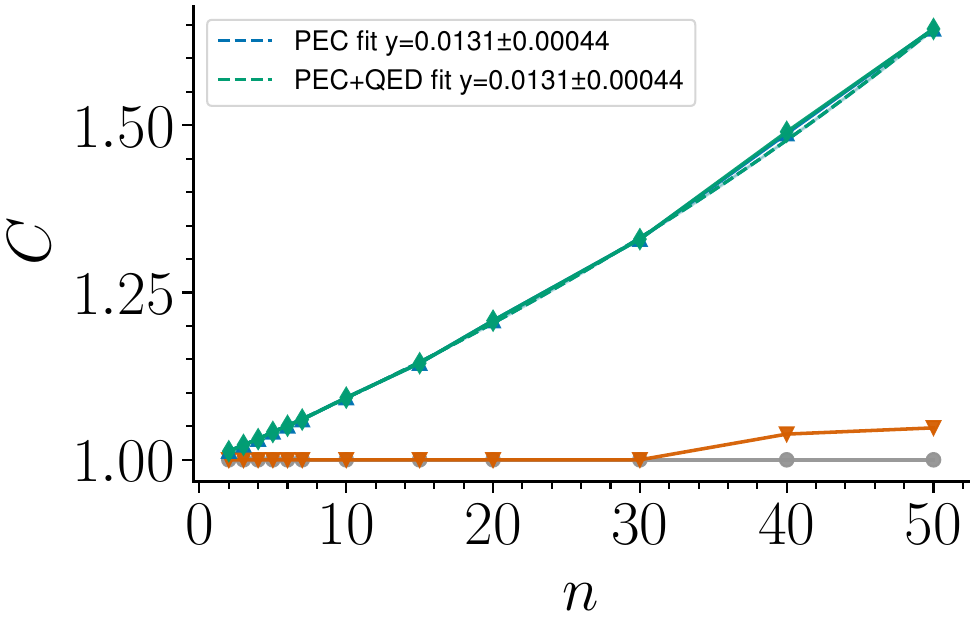} 
    \begin{picture}(0,0)
        \put(-160,108){\scriptsize (e)} 
    \end{picture}
    
    &
    \includegraphics[width=0.32\textwidth]{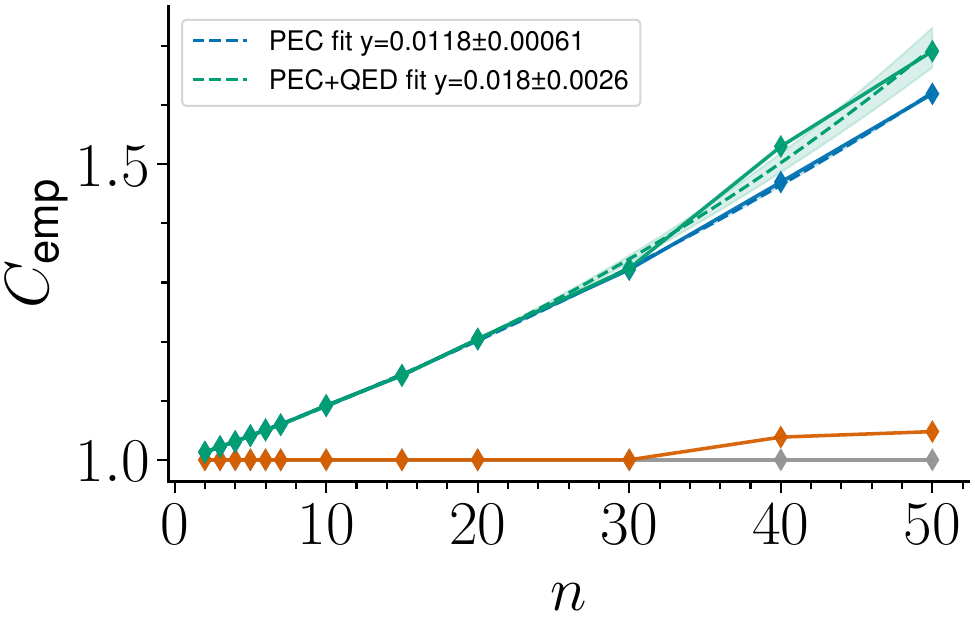}
    \begin{picture}(0,0)
        \put(-160,108){\scriptsize (f)} 
    \end{picture}
  \end{tabular}
  \caption{GHZ state preparation mitigation for linear and logarithmic depth circuit.
  The total square error (TSE) and mitigation costs between the output of a noise-free and noisy GHZ state preparation circuit with increasing GHZ state size $n$.
  Results obtained using stabilizer circuit simulations with: no mitigation ``Noisy" (grey), QED (orange), PEC (blue) and \qedpec (green), all simulations use $10^6$ shots.
  Results are shown for linear  (a), (b), (c)  and logarithmic depth (d), (e), (f) circuits.
  Mitigation costs are fit to an exponential $C(n) = x e^{y n} + z$ with $x + z \sim 1$.
  $C$ is the predicted sampling cost using PEC theory and empirical QED discard rate.
  $C_{\textrm{emp}}$ is the empirical sampling cost obtained by calculating the total number of discarded shots after applying PEC to the experiment counts.
  }
  \label{fig:ghz_res}
\end{figure*}

\textit{Stabilizer search outcome.}
To optimize stabilizer configurations for each preparation circuit we follow the procedure described in Sec.~\ref{sec:sym_selec} with the idling-noise extension applied, beginning with $\mathcal{S} = \mathbb{G}_{\scriptscriptstyle\textrm{GHZ}}$. 
The optimization involves searching over a subset of the $2^{n+1}$ sequences of generators and returns a sequence of stabilizers to measure, which is translated into a sequence of measurement circuits to implement in experiment.

For the linear-depth circuit, the optimization could not find a good configuration of sequentially applied stabilizer generator measurements.
This is because, even though each additional generator enabled the detection of more errors, therefore reducing the cost of applying PEC to undetectable errors [see Eq.~(\ref{eq:cost_pec_undet})], this reduction was not great enough to outweigh the additional overhead incurred by the measurement circuit mitigation [see Eq.~(\ref{eq:pec_cost_usym})].
As the $X^{\otimes n}$ check required $n$ CNOT gates, measuring this stabilizer introduces more noise than the target GHZ state preparation circuit, which requires $n - 1$ CNOT gates. 
Therefore, the $X^{\otimes n}$ check can be trivially excluded.
To investigate whether the optimization failure originates from either the low-weight $Z$-type checks or the qubit support of the generators, we expand the search to any weight-2 element of the GHZ stabilizer group.
We thus repeat the optimization with $\mathcal{S} = \mathbb{G}_{\scriptscriptstyle\textrm{GHZ}} \ \cup \ \{ Z_i Z_j : 1 \leq i < j \leq n\}$.
For $n \geq 10$, this finds low-weight non-local stabilizer configurations which detect sufficient errors for \qedpec to return a net sampling cost reduction compared to PEC.
For $10 \leq n \leq 40$ The optimization returns a stabilizer of the form
\begin{align} \label{eq:ghz_lin_ss}
    Z_0 I_1 I_2 \cdots I_{n-3} I_{n-2} Z_{n-1},\\ 
    I_0 Z_1 I_2 \cdots I_{n-3} Z_{n-2} I_{n-1},\\
    Z_0 I_1 I_2 \cdots I_{n-3} Z_{n-2} I_{n-1},\\
    I_0 Z_1 I_2 \cdots I_{n-3}I_{n-2} Z_{n-1},
    \label{eq:ghz_lin_ss_final}
\end{align}
where all four solutions achieve the same objective value, see Tab.~\ref{tab:ghz_lin_scores}.
At $n = 50$, the set $\mathcal{T}^* = \{ Z_i Z_j \ | \ (i, j) \in  \{(0, 49), (24, 25), (23, 26), (22, 27), (21, 28), (20, 29) \}$ is returned.
For the logarithmic-depth circuit,
we can only find a stabilizer configuration with a net benefit for $n \geq 15$.
Unlike the linear-depth example, we find configurations using $\mathcal{S} = \mathbb{G}_{\scriptscriptstyle\textrm{GHZ}}$. Crucially, a lower cost can be reached by searching with $\mathcal{S} = \mathbb{G}_{\scriptscriptstyle\textrm{GHZ}} \ \cup \ \{ Z_i Z_j : 1 \leq i < j \leq n\}$.
We include these measurement configurations and optimization scores in Tab. \ref{tab:ghz}.

\begin{table}[]
    \centering
    \begin{tabular}{c | c | c | c}
      $n$  &  $ \left(\gamma_{\scriptscriptstyle \text{undet}} \right)^2 \times $ & $\left(\gamma_{\scriptscriptstyle \text{undet}} \right)^2 \times $  & $\gamma^2$ \\ 
       & $\ \ C_{\scriptscriptstyle \textrm{SYM}}(\mathcal{T}^*)$ & $C_{\scriptscriptstyle \textrm{SYM}}(\mathbb{G}_{\scriptscriptstyle\textrm{GHZ}})$ & 
      \\\hline 
     10  & 1.050 & 1.267 & 1.054 \\
     15  & 1.08 & 1.52 & 1.10 \\
     20  & 1.12 & 1.91 & 1.16 \\
     30  & 1.23 & 3.39 & 1.33 \\
     40  & 1.38 & 7.13 & 1.59 \\
     50  & 1.58 & 17.7 & 1.97
    \end{tabular}
    \caption{Optimization scores for stabilizers $\mathcal{T}^*$ in Eqs.~(\ref{eq:ghz_lin_ss}-\ref{eq:ghz_lin_ss_final}) for procedure  in Sec.~\ref{sec:sym_selec_a} applied to a GHZ state preparation circuit with depth growing linearly in size $n$ (excluding terminating computational basis measurements). 
    Optimization scores are given by the sampling cost of applying PEC to undetectable errors $\left(\gamma_{\scriptscriptstyle \text{undet}} \right)^2 $ and stabilizer measurement circuit errors $C_{\scriptscriptstyle \textrm{SYM}}(\mathcal{C})$.
    When $\mathcal{C} = \mathbb{G}_{\scriptscriptstyle\textrm{GHZ}}$ all stabilizer generators are used. $\gamma^2$ is the cost of applying PEC to all state preparation errors.
    }
    \label{tab:ghz_lin_scores}
\end{table}

\textit{Mitigated distribution accuracy.}
We test the selected stabilizers by constructing mitigated output distributions from noisy circuit simulations. We also estimate the full sampling cost of \qedpec using empirical estimates of $p_{\nd}^{\text{tot}}$.

At small $n < 10$, \qedpec is executed without performing QED, giving equivalent TSE to PEC within finite-shot noise and equivalent sampling overheads, see Fig.~\ref{fig:ghz_res}(a)-(c).
At increasing $n \geq 10$, the smaller sampling overhead of \qedpec manifests as a reduction in TSE compared to PEC, growing to just under an order of magnitude at $n = 50$.
QED is performed with the same stabilizers used for \qedpec and improves over unmitigated results by half an order of magnitude at $n = 10$ up to an order of magnitude at $n = 50$.
This modest performance is because QED using Eq.~(\ref{eq:ghz_lin_ss}) for a stabilizer is not sufficient to detect any single qubit error, so a large fraction of errors pass undetected.
The sampling cost scalings are well fit by exponential curves, see Fig.~\ref{fig:ghz_res}(b) and (c), with the reduction of \qedpec increasing with $n$ after the initial crossover $n$ is passed. 
In Fig.~\ref{fig:ghz_res}(d), we see that for logarithmic-depth circuits, the stabilizers selected by our optimization give an advantage for \qedpec over PEC of around half an order of magnitude in TSE.
However, for $n \geq 40$, in Fig.~\ref{fig:ghz_res}(e), we see that the cost of \qedpec with post-selection taken into account is approximately equal to the cost of PEC.
As in Fig.~\ref{fig:ghz_res}(c), we find the empirical sampling cost in Fig.~\ref{fig:ghz_res}(f) is larger than the predicted cost in Fig.~\ref{fig:ghz_res}(e). 

\begin{figure*}
    \centering
    \begin{tabular}{c c}(a)
       \includegraphics[width=0.45\linewidth]{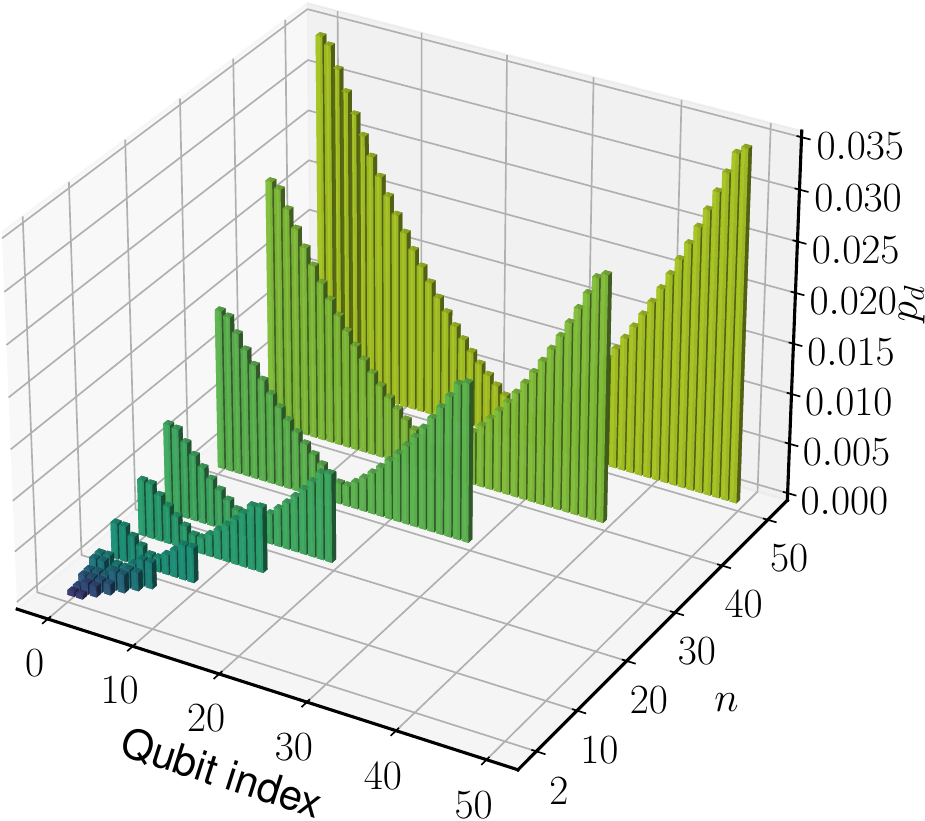}   
       &
       (b)
       \includegraphics[width=0.45\linewidth]{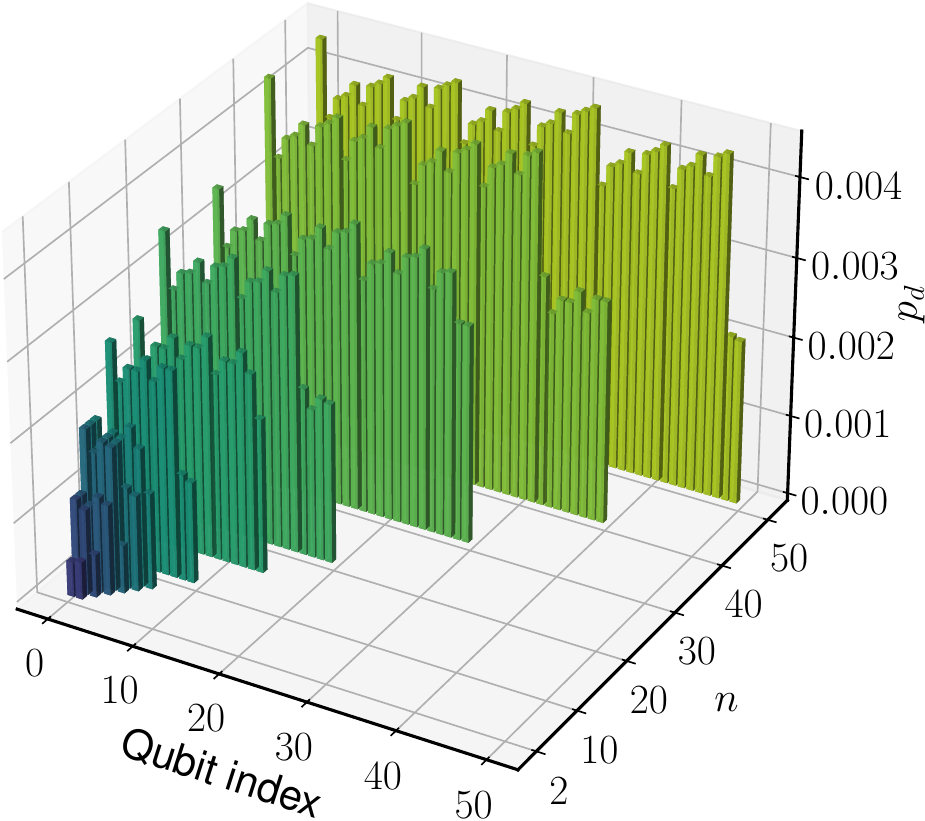} 
    \end{tabular}
    \caption{Distribution of detectable error probability $p_d$ across qubits after propagation and composition of noise channels through $n$-qubit GHZ state preparation circuits with (a) linear and (b) logarithmic depth. Propagated errors are classified as detectable if they anticommute with at least 1 $Z$-type GHZ state stabilizer generator. Colours correspond to different values of $n$.}
    \label{fig:ghz_error_prop}
\end{figure*}

This performance discrepancy stems from the fact that the logarithmic-depth circuit is shallower, but also from how errors propagate through each circuit. This is seen by propagating all errors through each preparation circuit, not including stabilizer measurements, and observing how error probabilities accumulate on different qubits. 
For the linear-depth circuit, detectable errors accumulate with higher probability on a smaller set of qubits increasingly as $n$ grows, see Fig.~\ref{fig:ghz_error_prop}(a). 
For the logarithmic-depth circuit, detectable error probability is spread more uniformly across qubits, see Fig.~\ref{fig:ghz_error_prop}(b). 
Therefore, the small number of low-weight stabilizers we use, as in Eq.~(\ref{eq:ghz_lin_ss}-\ref{eq:ghz_lin_ss_final}) 
and Tab.~\ref{tab:ghz}, are capable of detecting a higher proportion of errors for the linear-depth circuit than the logarithmic-depth circuit.

Without idling noise, we observe qualitatively similar results for the linear-depth circuit. Stabilizers of the same form as Eq.~(\ref{eq:ghz_lin_ss}-\ref{eq:ghz_lin_ss_final}) are selected.
For the logarithmic-depth circuits, we observe a smaller $\textrm{TSE}$ for \qedpec than PEC for $n \geq 30$ and a smaller predicted sampling cost for \qedpec and a larger empirical cost. In contrast to the linear-depth case, higher weight stabilizers are selected.
We include additional figures and selected stabilizers for simulations without idling noise in App.~\ref{app_ghz}. 

\begin{table*}[]
    \centering
    \begin{tabular}{c | c | c | c | c}
      $n$  & $\mathcal{T}^*$ & $ \left(\gamma_{\scriptscriptstyle \text{undet}} \right)^2 \times $ & $\left(\gamma_{\scriptscriptstyle \text{undet}} \right)^2 \times $  & $\gamma^2$ \\ 
       &  & $\ \ C_{\scriptscriptstyle \textrm{SYM}}(\mathcal{C})$ & $C_{\scriptscriptstyle \textrm{SYM}}(\mathbb{G}_{\scriptscriptstyle\textrm{GHZ}})$ & 
      \\\hline 
     40 & $Z_{15} Z_{31}$, $Z_{7} Z_{39}$, $Z_{23} Z_{27}$, $Z_{3} Z_{11}$, $Z_{19} Z_{35}$ & 1.265 & 6.916 & 1.267\\
     50 & $Z_{31} Z_{47}$, $Z_{15} Z_{23}$, $Z_{7} Z_{39}$, $Z_{11} Z_{19}$, $Z_{3} Z_{27}$, $Z_{35} Z_{43}$ & 1.34 & 16.8 & 1.35
    \end{tabular}
    \caption{Stabilizer configurations $\mathcal{T}^*$ selected by the optimization in Sec.~\ref{sec:sym_selec_a} for a GHZ state preparation circuit with depth growing logarithmically in size $n$ (excluding terminating computational basis measurements). 
    Optimization scores are given by the sampling cost of applying PEC to undetectable errors $\left(\gamma_{\scriptscriptstyle \text{undet}} \right)^2 $ and stabilizer measurement circuit errors $C_{\scriptscriptstyle \textrm{SYM}}(\mathcal{C})$.
    When $\mathcal{C} = \mathbb{G}_{\scriptscriptstyle\textrm{GHZ}}$ all stabilizer generators are used. $\gamma^2$ is the cost of applying PEC to all state preparation errors.
    }
    \label{tab:ghz}
\end{table*}

\subsection{Fermi-Hubbard model time dynamics} \label{sec:fh}
We simulate the spinless Fermi-Hubbard (FH) model on a square lattice with $m$ sites, containing $N$ fermions. This model is described by the Hamiltonian
\begin{multline} \label{eq:fh_sless}
H_{S} = - \tau \sum_{\langle i, j \rangle \in E} \left( a^\dagger_{i} a_{j} + a_{i} a^\dagger_{j,} \right) +  u \sum_{\langle i, j \rangle \in E} n_{i} n_{j},
\end{multline}
where $a^\dagger_{i}$ ($a_{i}$) creates (annihilates) a spinless fermion at site $i$ and $n_{i} = a^\dagger_{i} a_{i}$. The operators satisfy $\{a_{i}, a^\dagger_{j} \} = \delta_{ij}$, $\{a_{i}, a_{j} \} = 0$, $\{a^\dagger_{i}, a^\dagger_{j} \} = 0$ for $\{A, B\} = AB + BA$. 
$u$ quantifies the interaction strength between sites $i$ and $j$. 
$\tau$ quantifies the hopping rate between sites.
We work in units where $\hbar = 1$.
Fermions are placed on the vertices $V$ of a graph $G = (V, E)$ where interactions are specified by the edge set $E$.
The spinless FH model is among the simplest interacting fermionic lattice models and exhibits interaction-driven quantum phase transitions, making it a useful benchmark for near-term quantum simulation.

To simulate time evolution of a fermionic system using qubits we map creation and annihilation operators to Pauli operators that reproduce fermionic statistics. 
There are several different fermion to qubit encodings, each with different strengths and weaknesses.
The generalized superfast encoding (GSE)~\cite{setiaSuperfastEncodingsFermionic2019}  maps $m$ interacting sites on a degree-$d$ graph onto $\mathcal{O}(md)$ qubits acted on by weight $\mathcal{O}(d)$ Pauli string operators. 
These Pauli operators are grouped in vertex and edge operators which represent on-site and interaction terms in Eq.~(\ref{eq:fh_sless}).
To satisfy the canonical fermionic anti-commutation relations, products of edge operators around closed loops in the graph must equal the identity. In the GSE this condition is satisfied by forcing fermionic states to lie in a simultaneous +1 eigenspace of all loop operators during state preparation. Therefore, loop operators form a quantum stabilizer code which both ensures that states obey correct fermionic statistics and enables error correction or detection.

We use the GSE proposed in~\cite{haggeErrorMitigationError2023} which enables detection of single qubit errors for the spinless FH model on a square lattice.
The encoding is constructed by placing two qubits at each fermion site and replacing each term in \eqn{\ref{eq:fh_sless}} with vertex and edge operators acting on and between these groups of qubits.
Each vertex operator is 
\begin{align} \label{eq:vop}
B_j=I-2 a^\dagger_{j} a_{j},    
\end{align}
and each edge operator is
\begin{align}
A_{j,k}=-\mathrm{i}(a_{j}+a^\dagger_{j,})(a_{k}+a^\dagger_{k}).
\end{align}
The qubit encoded Hamiltonian is 
\begin{multline} \label{eq:gse_ham}
     \widetilde{H}_S = \frac{\mathrm{i} \tau}{2} \sum_{\langle j, k \rangle \in E} B_j A_{(j, k)} + A_{(j, k)} B_k + \\
     \frac{u}{4} \sum_{\langle j, k \rangle \in E} (1 - B_j) (1 - B_k),
\end{multline}
with qubit encoded vertex $B_j = - Z_{j, 0} Y_{j, 1}$ and edge operators $A_{(j, k)} \in \{\pm I_{j, 0} X_{j, 1} Y_{k, 0} Y_{k, 1}, \pm I_{j, 0} Z_{j, 1} X_{k, 0} Y_{k, 1} \} $ for Pauli operators $P_{j,i}$ where $P \in \{I, X, Y, Z \}$ acting on the $i$th qubit at the $j$th fermion site.
The edge operator sign $\epsilon_{jk} = \pm 1$ is set by the edge orientation, such that $\epsilon_{kj} = - \epsilon_{jk}$ and edges can be vertical $A_{(j, k)} = \pm I_{j, 0} X_{j, 1} Y_{k, 0} Y_{k, 1}$ or horizontal $A_{(j, k)} = \pm I_{j, 0} Z_{j, 1} X_{k, 0} Y_{k, 1}$.
We give an explicit construction of \eqn{\ref{eq:gse_ham}} and the stabilizing loop operators for a $2 \times 2$ lattice in App.~\ref{app:gse_2x2}.

As an example we mitigate observables after time evolution under Eq.~(\ref{eq:gse_ham}).
For an observable $O$, written as a sum of products of lattice operators, evolution time $t$ and initial state $\ket{\psi_0}$, we estimate $\langle O(t)\rangle=~\bra{\psi_0}e^{\mathrm{i} t \widetilde{H}_S}Oe^{-\mathrm{i} t \widetilde{H}_S}\ket{\psi_0}$.
To approximate $e^{-\mathrm{i} \widetilde{H}_S t}$ we use $x$ 1st-order Trotter steps 
\begin{align} \label{eq:trot}
U(\Delta t, x, \tau, u) &=
\Bigl[ \prod_{\langle j,k \rangle \in E} 
  e^{-\mathrm{i} \Delta t \frac{\tau}{2} (\mathrm{i} B_j A_{(j,k)})} 
\notag\\
&\times
\prod_{\langle j,k \rangle \in E} 
  e^{-\mathrm{i} \Delta t \frac{\tau}{2} (\mathrm{i} A_{(j,k)} B_k)} \notag\\
&\times
\prod_{j \in V} 
  e^{\mathrm{i} \Delta t\frac{u}{2} B_j}
\prod_{\langle j,k \rangle \in E} 
  e^{-\mathrm{i} \Delta t B_j B_k \frac{u}{4}} \Bigr]^x,
\end{align}
where each step is evolved for time $\Delta t = t/x$ and we omit the global phase factor.

To implement Eq.~(\ref{eq:trot}) on a digital quantum computer we must decompose each term into gates from a finite basis set.
We use a Clifford and $R_P(\theta) = \exp(-\mathrm{i}\theta P)$ basis, as each term in Eq.~(\ref{eq:trot}) can then be implemented fault-tolerantly with the prescription in Sec.~5 of~\cite{haggeErrorMitigationError2023}.

\begin{figure*}[ht] 
\centering 
    \includegraphics[width=0.95\textwidth]{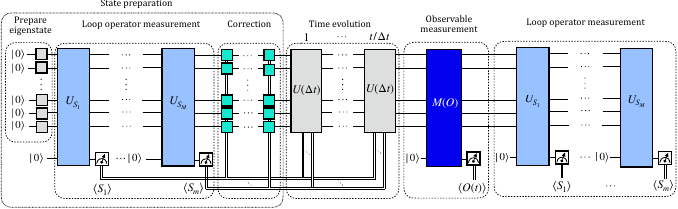}

\caption{Circuits used to study time evolution under the Fermi-Hubbard model with a generalized superfast fermion-to-qubit encoding. 
State preparation: prepare an eigenstate of both vertex operators and stabilizing loop operators by applying single-qubit Clifford gates, measuring all loop operator expected values $\{ \langle S_i \rangle\}_{i=1}^m$ and applying corrective operations based on the measurement outcomes, see App.~\ref{app:prep} for more detail.
Time evolution: implement approximate evolution to time $t$ by applying $t / \Delta t$ repeats of a Trotter step $U(\Delta t)$ evolved to time $\Delta t$, see Eq.~(\ref{eq:trot}). Due to vertex operator sign redefinition during state preparation, each Trotter step is classically controlled on loop-operator measurement outcomes.
If using QED-based QEM, measure loop operators again, use results for post-selection.
Observable measurement: measure lattice operator e.g. $B_j$ or $B_j A_{jk}$ using ancilla qubit and circuit $M(O)$, to give $\langle O(t) \rangle$.
Circuits for time evolution, loop operator and observable measurement circuits are shown in App.~\ref{app:fh_circ}.}
\label{fig:fh_circ}
\end{figure*}

We prepare the initial state of $N$-fermions $\ket{\psi_0}$ by adapting the definite-occupancy state preparation method proposed in Sec.~3.3 of~\cite{haggeErrorMitigationError2023}. We prepare a state with a fixed fermion number,  measure each loop operator, then using classical feedforward, apply corrections to ensure the state is a +1 eigenstate of all loop operators.  
As loop operator measurements commute with the vertex operator measurements, but the correction operations don't, we then redefine the sign of each lattice vertex operator using feedforward to maintain the correct fermion number. 
Fig.~\ref{fig:fh_circ} shows a schematic of the full circuit.
State preparation is followed by the circuits for time evolution, lattice operator measurement and finally stabilizer measurement circuits for techniques using QED.
We use the noise model described in Sec.~\ref{sec:ghz}.
We use a single ancilla qubit to perform stabilizer and lattice operator measurements, applying a noisy qubit reset after each measurement except the final one.
We provide more detail on state preparation in App.~\ref{app:prep}
and describe the time evolution and measurement circuits in more detail in App.~\ref{app:fh_circ}.

We mitigate increasing-weight products of vertex operators $B_i$, which arise naturally when evaluating multi-site correlation functions, as well as products of vertex and edge operators $B_j A_{jk}$ and $A_{jk} B_k$, which correspond directly to the hopping terms of the Hamiltonian in Eq.~(\ref{eq:gse_ham}). Together they span the observable classes whose accurate estimation is required for near-term fermionic simulation in the GSE.

Phases in the Fermi-Hubbard model, such as a Mott Insulator or metallic phases are characterized by a two-point or radial charge correlator~\cite{stanisicObservingGroundstateProperties2021}
\begin{align} \label{eq:norm_cc}
    G_{ij} = \frac{\langle n_i n_j \rangle - \langle n_i \rangle \langle n_j \rangle}{\langle n_i^2 \rangle - \langle n_i \rangle^2},
\end{align}
and
\begin{align} \label{eq:radial_cc}
    G(r) = \frac{1}{L^2} \sum_{i \in V} \frac{1}{N_{r, i}} \sum_{j \in S_r(i)} \langle (n_i - \overline{n})(n_j - \overline{n}) \rangle,
\end{align}
where $\overline{n} = N / m$ is the fermion density, $L = \sqrt{m}$, is the square lattice side-length, $S_r(i)$ is the set of vertices distance $r$ away from $i$ on the lattice and $N_{r, i} = \vert S_r(i) \vert$~\cite{stokesPhasesTwodimensionalSpinless2020}. Since such quantities are of general interest, we characterize how well we can mitigate them.
The $G_{ij}$ and $G(r)$ are written in terms of vertex operators using $ B_i  = 2 n_i - 1$.

Vertex operator measurements can be performed fault tolerantly in the GSE. Therefore, detectable errors remain detectable~\cite{haggeErrorMitigationError2023} such that QED can be done after observable measurement. 
For products of vertex operators and edge operators, observable measurement is not fault-tolerant without additional resources e.g. flag qubits, see App.~\ref{app:fh_circ}. Therefore, in this case we perform QED before observable measurement. 

We choose stabilizer configurations by performing the optimization described in Eq.~({\ref{eq:opt_fd}}).
We find that for $m = 2,3, 4$, the optimization identifies the full set of available loop operators as the configuration which minimizes the mitigation sampling cost. 
Crucially, for $m = 5,6$ the optimization identifies partial stabilizer configurations with a lower sampling cost. 
For $m = 5$ and $6$ we find that measuring only 83\% and 78\% of the respective stabilizers is beneficial. 
When computing sampling costs for increasing $m$, we include results for QED and \qedpec with the full and partial stabilizer configurations.

\begin{figure}[ht]
  \centering    \includegraphics[width=0.49\textwidth]{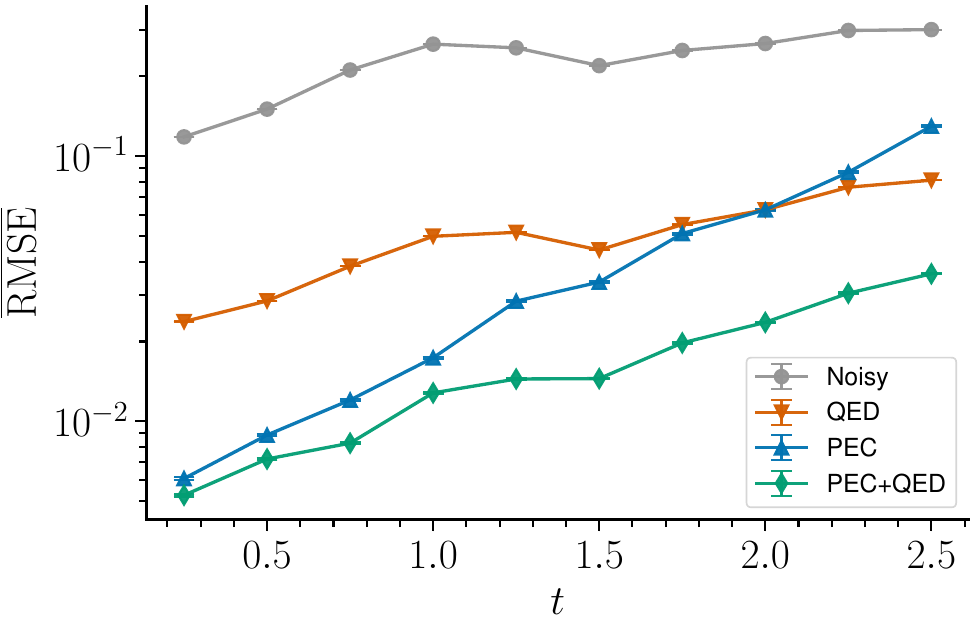} 
    
  \caption{Mitigation accuracy for Hamiltonian and vertex operator observables in simulating Fermi-Hubbard model time dynamics using an error detecting GSE. 
  The estimated root mean squared error $\overline{\textrm{RMSE}}$ averaged over 23 observables, including each term in Eq.~(\ref{eq:gse_ham}) and weight-3 and -4 vertex operator products, for a 2$\times$2 lattice.
  Results are compared for no mitigation ``Noisy" (grey circles), PEC (blue triangles) and symmetry-based approaches: QED (orange triangle down) and \qedpec (green diamonds).
  Error bars show the standard error.}
  \label{fig:fh_res_main_error}
\end{figure}

\begin{figure*}[ht]
  \centering
  \begin{tabular}{@{}cc@{}}
    \includegraphics[width=0.49\textwidth]{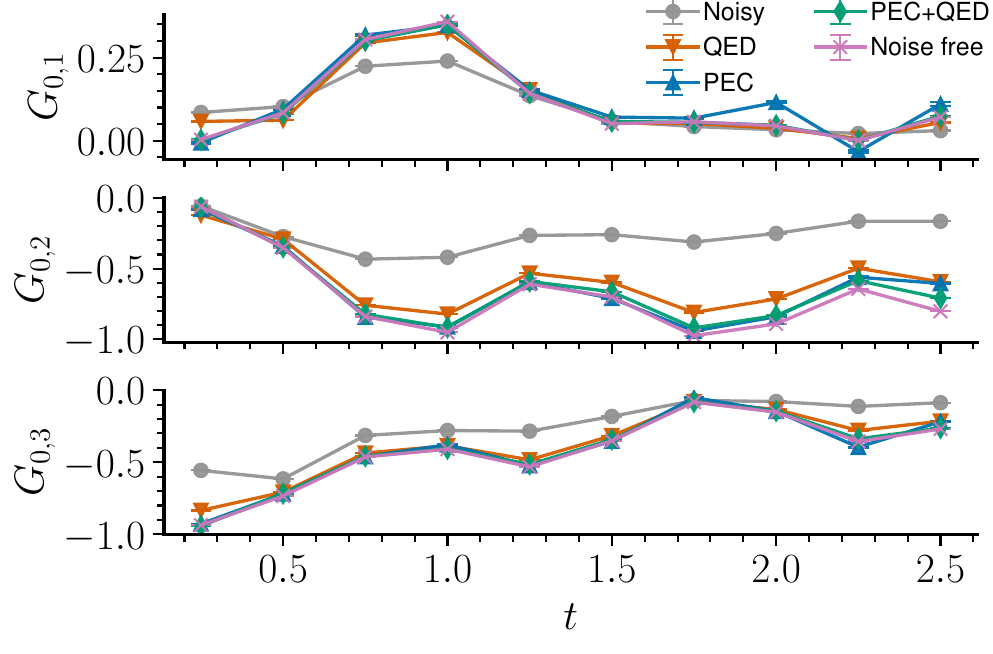} 
    \begin{picture}(0,0)
        \put(-230,160){\scriptsize (a)} 
    \end{picture}   
    &
    \includegraphics[width=0.49\textwidth]{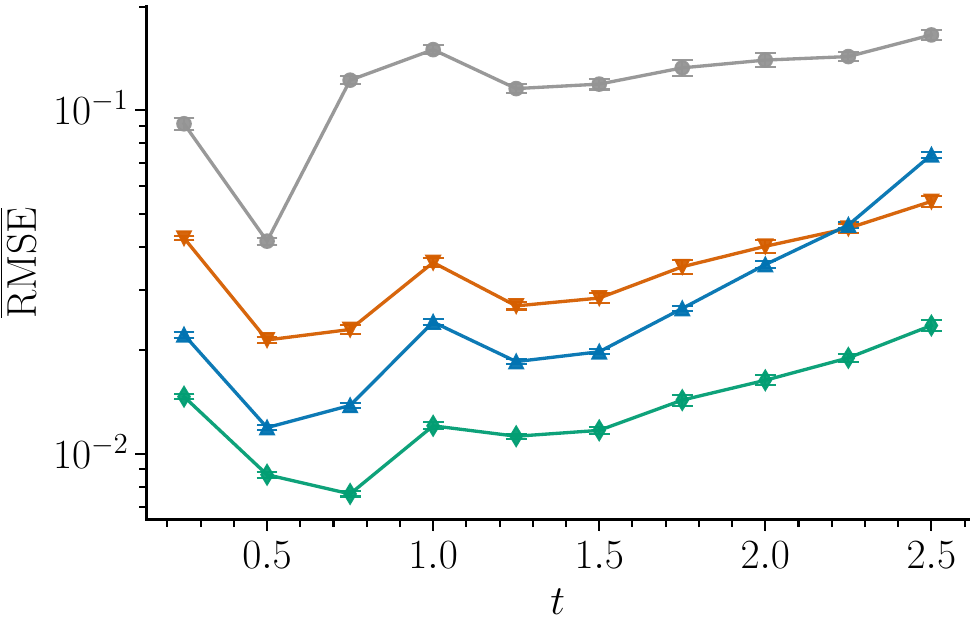} 
    \begin{picture}(0,0)
        \put(-230,160){\scriptsize (b)} 
    \end{picture}

  \end{tabular}
  \caption{Mitigation accuracy for a two-point charge correlator during simulated Fermi-Hubbard model time dynamics using an error detecting Generalized Superfast encoding. 
  (a) Correlator $G_{0j}$ between the 0th and $j$th sites for estimation with no noise: ``Noise free" (pink crosses), no mitigation: ``Noisy" (grey circles), PEC (blue triangles) and symmetry-based approaches: QED (orange triangle down) and \qedpec (green diamonds).
  (b) The estimated root mean squared error $\overline{\textrm{RMSE}}$ of the correlator estimator averaged over $j$.
  Error bars show the standard error.}
  
  \label{fig:norm_cc}
\end{figure*}
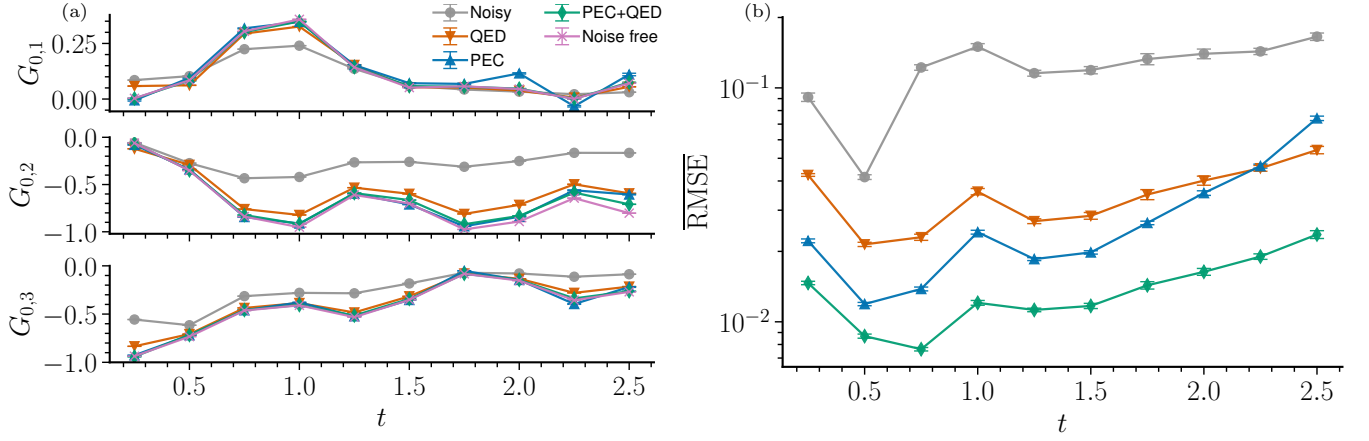

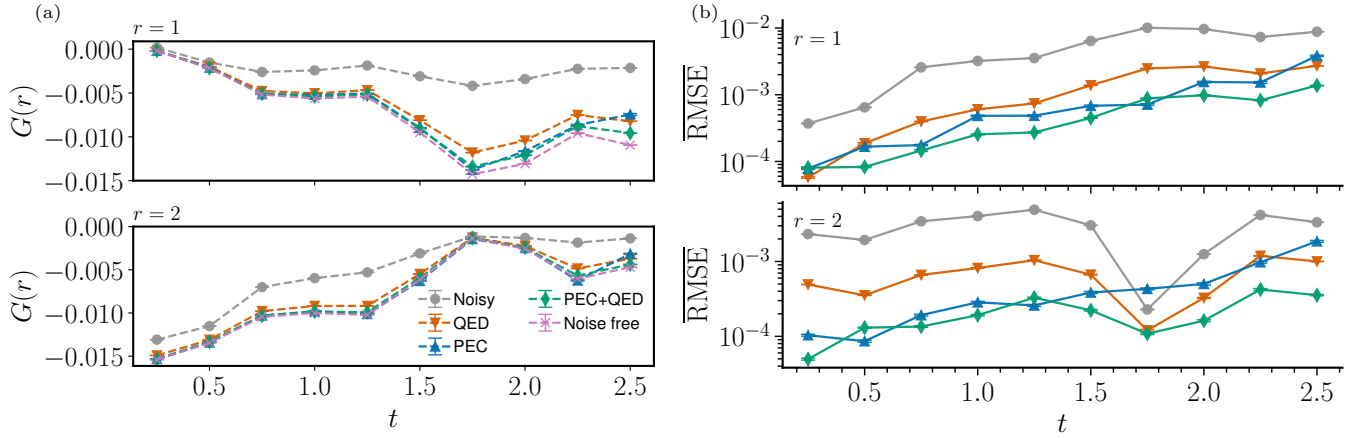
\begin{figure*}[ht]
  \centering
  \begin{tabular}{@{}cc@{}}
    \includegraphics[width=0.49\textwidth]{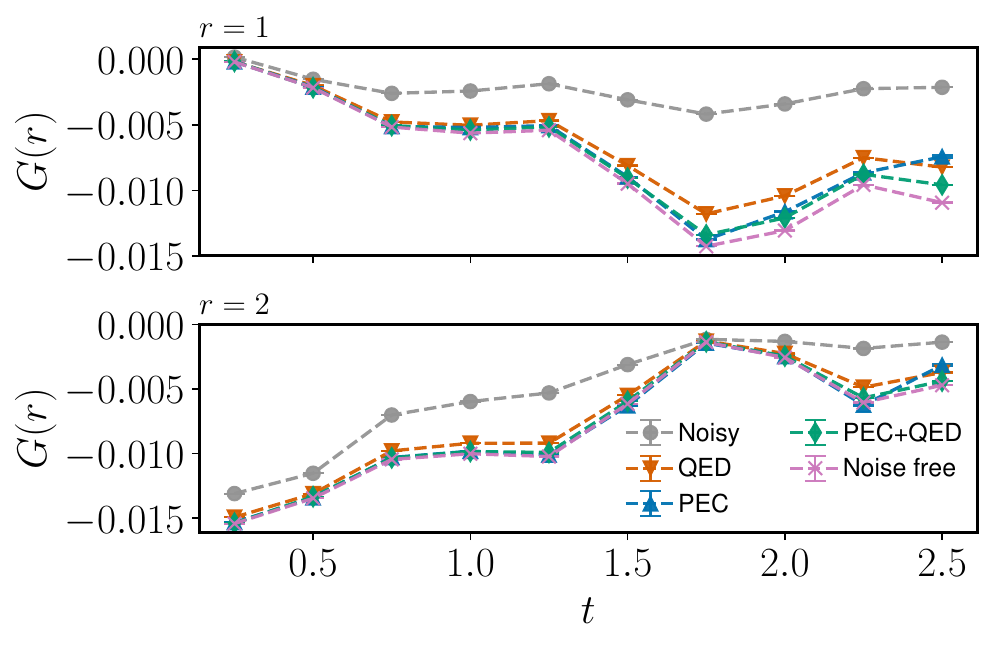} 
    \begin{picture}(0,0)
        \put(-240,160){\scriptsize (a)} 
    \end{picture}
    
    &
    \includegraphics[width=0.49\textwidth]{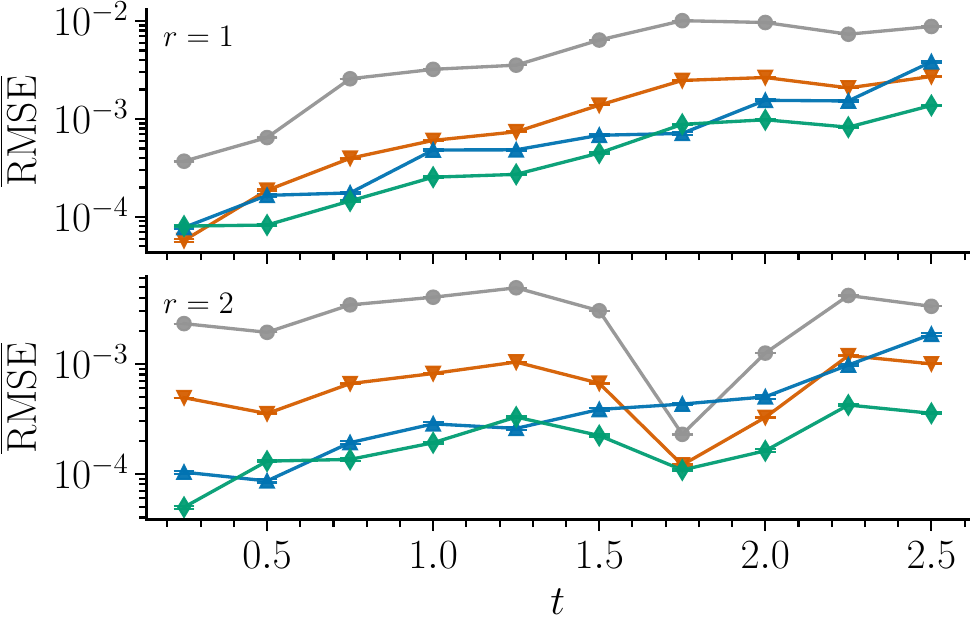} 
    \begin{picture}(0,0)
        \put(-250,160){\scriptsize (b)} 
    \end{picture} 
  \end{tabular}
  \caption{Mitigation accuracy for a radial charge correlator during simulated Fermi-Hubbard model time dynamics using an error detecting Generalized Superfast encoding. 
  (a) Correlator $G(r)$ at distances $r = 1$ (top) and $r = 2$ (bottom) for estimation with no noise: ``Noise free" (pink crosses), no mitigation: ``Noisy" (grey circles), PEC (blue triangles) and symmetry-based approaches: QED (orange triangle down) and \qedpec (green diamonds).
  (b) The estimated root mean squared error $\overline{\textrm{RMSE}}$ of the correlator estimator for $r = 1,2$.
  Error bars show the standard error.}  
  \label{fig:radial_cc}
\end{figure*}

\begin{figure}[ht]
  \centering
    \includegraphics[width=0.49\textwidth]{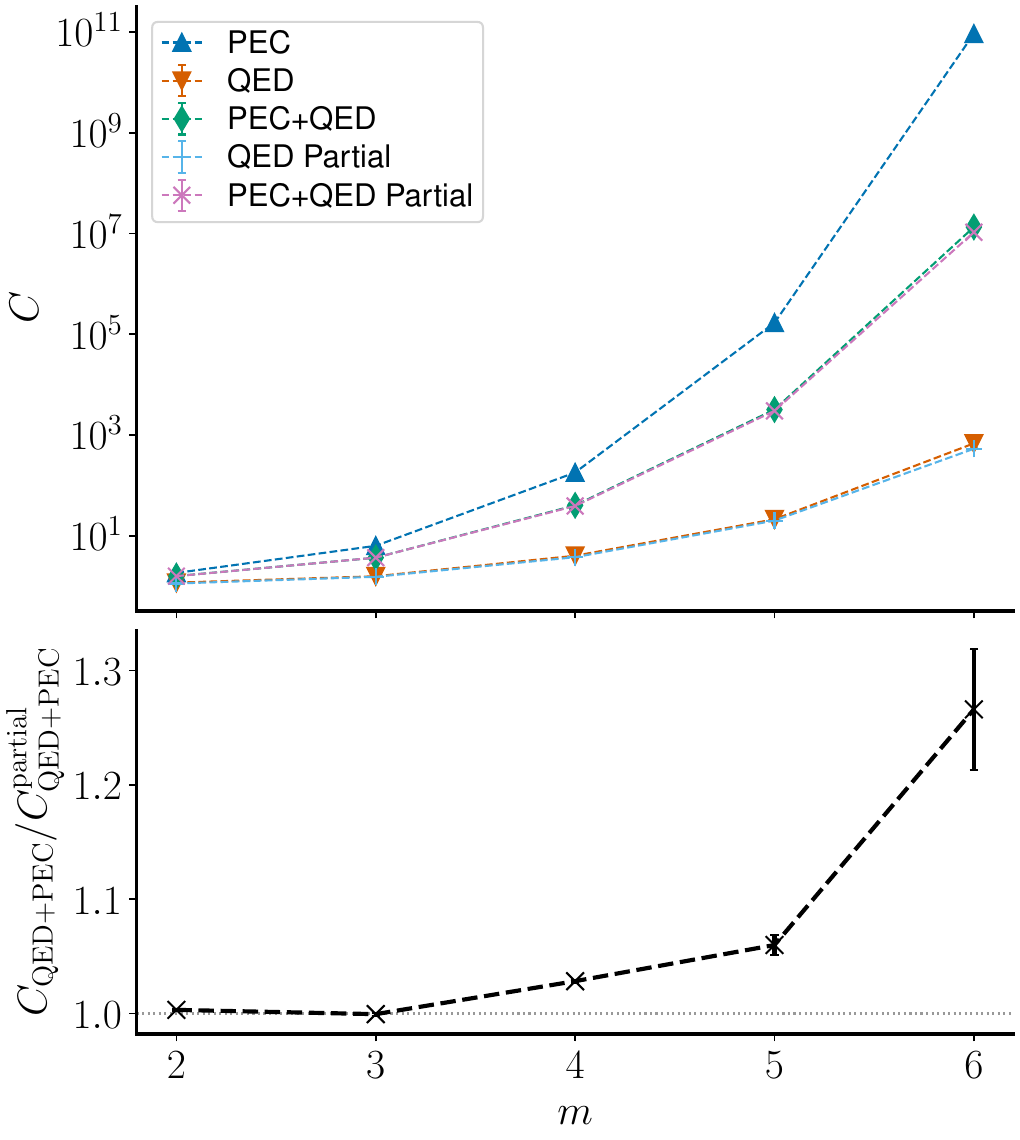}
    \begin{picture}(0,0)
        \put(-120,290){\scriptsize (a)} 
    \end{picture} 
    \begin{picture}(0,0)
        \put(-125,140){\scriptsize (b)} 
    \end{picture} 
  \caption{
  Sampling costs for simulating Fermi-Hubbard model time dynamics using an error detecting Generalized Superfast encoding.
  (a) Sampling overhead $C$ for one 1st-order Trotter step simulation at $t = \pi$ for an $m \times m$ lattice. 
  Results are compared for PEC (blue triangles) and symmetry-based approaches: QED (orange triangle down) and \qedpec (green diamonds), which were also run with reduced symmetry measurement configurations QED Partial (light-blue plus) and \qedpec Partial (pink cross), obtained using Eq.~(\ref{eq:opt_fd}).
  (b) The ratio of the sampling costs of \qedpec and \qedpec Partial.
  Error bars from numerical estimates of post-selection rates show the standard error. 
  }
  \label{fig:fh_res_main_cost}
\end{figure}

\textit{Observable mitigation for a 2 $\times$ 2 model.}
Unless otherwise stated, subsequent simulations use the following parameters.
We consider the FH model on a $2\times2$ lattice evolved from time $t = 0.250$ to 2.5 in steps of $\Delta t = 0.125$. The fermion density is $\overline{n} =  0.50$, the hopping rate $\tau = 1$ and interaction strength $u = 4$. For each time step we estimate each term in Eq.~(\ref{eq:gse_ham}) and weight-3 and -4 vertex operator products both with noise, for $p = 10^{-3}$ and $N_S = 10^5$ and without noise, for $p = 0$ and $N_S = 10^6$.

As shown in Fig.~\ref{fig:fh_res_main_error}, we observe that all mitigation techniques perform substantially better than the noisy results, giving around an order of magnitude reduction in the root mean square error averaged over bootstraps 
\begin{align}
   \overline{\textrm{RMSE}}(t) = \frac{1}{N_B} \sum_{i = 1}^{N_B} \sqrt{\frac{1}{N_O} \sum_{j = 1}^{N_O}  \left( \tilde{O}^{\textrm{nf}}_j(t) - \overline{O}^{(i)}_j(t)\right)^2}.
\end{align}
Here, the number of bootstraps and measured observables are  $N_B=100$ and $N_O=23$, respectively. 
Furthermore, $\overline{O}^{(i)}_j(t)$ is the sample estimate of the $j$th observable at the $i$th bootstrap and its corresponding noise-free reference is $\tilde{O}^{\textrm{nf}}_j(t) = (\sum_{i = 1}^{N_B} \overline{O}^{\textrm{nf}, (i)}_j(t)) / N_B$. We obtain $N_B$ expectation value estimates by parametrically resampling each observed measurement count distribution.

At $0.25 \leq t \leq 1.25$ the PEC based techniques have a lower $\overline{\textrm{RMSE}}$ than QED alone, however, as $t$ and therefore circuit depth increases, the variance of the PEC estimator grows so that from $t = 2.0$, the $\overline{\textrm{RMSE}}$ for PEC is comparable or greater to that of QED, see Fig.~\ref{fig:fh_res_main_error}. On the other hand, the symmetry reduced variance of the \qedpec estimator means it has the smallest $\overline{\textrm{RMSE}}$ for all $t$.

Fig.~\ref{fig:norm_cc}(a) shows the noise-free and mitigated estimates of the normalized two-point charge correlators $G_{i,j}$ defined in Eq.~(\ref{eq:norm_cc}) with $i = 0$ and $j = $1,2,3. The $\overline{\textrm{RMSE}}$ taken over $j$, i.e., 
\begin{align}
   \overline{\textrm{RMSE}}(t) = \frac{1}{N_B} \sum_{i = 1}^{N_B} \sqrt{\frac{1}{3} \sum_{j = 1}^{3}  \left( \tilde{G}^{\textrm{nf}}_{0,j}(t) - \overline{G}^{(i)}_{0,j}(t)\right)^2},
\end{align}
is shown in Fig.~\ref{fig:norm_cc}(b).
The advantage of applying mitigation over no mitigation is again substantial. In this case PEC achieves a small, approximately constant reduction in error over QED up to $t = 1.5$, after which its performance degrades leading to an increase in error at $t = 2.5$.
In comparison, \qedpec achieves the smallest $\overline{\textrm{RMSE}}$ by a similar magnitude to that in Fig.~\ref{fig:fh_res_main_error}, with a rate of increasing error appearing to more closely match that of QED than PEC for $t > 1$.
The initially large error for all techniques can be attributed to $G_{0,1}$, $G_{0,2}$ having noise-free values close to 0 and therefore harder to resolve in the presence of hardware and finite-shot noise.

Fig.~\ref{fig:radial_cc}(a) shows the noise-free and mitigated values of the radial charge correlator $G(r)$ defined in Eq.~(\ref{eq:radial_cc}) and the $\overline{\textrm{RMSE}}$ for each mitigated estimator in (b), i.e., 
\begin{align}
   \overline{\textrm{RMSE}}(t) = \frac{1}{N_B} \sum_{i = 1}^{N_B} \sqrt{ \left( \tilde{G}^{\textrm{nf}}(r, t) - \overline{G}^{(i)}(r, t)\right)^2}.
\end{align}
The unmitigated noisy estimator has approximately an order of magnitude higher $\overline{\textrm{RMSE}}$ for each $t$ with the exception of $t = 1.75$ where the decay of the noisy expectation towards 0 coincides with the noise-free expectation. 
All techniques achieve 1 to 2 orders of magnitude lower $\overline{\textrm{RMSE}}$ than for the previous correlator or average over lattice operators. Additionally the performance difference between each technique is much smaller, with PEC and \qedpec generally performing comparably except for $r = 2$ and $t > 1.5$.

\textit{Sampling cost for larger lattice sizes.}
In Fig.~\ref{fig:fh_res_main_cost} we show how the QEM sampling overheads for each approach scale with increasing lattice size $m$. 
The overheads grow faster than exponentially in $m$ for each technique. 
The difference in costs between PEC and \qedpec grows with $m$, suggesting improved utility at larger lattice sizes. 
Measuring partial stabilizer configurations gives a reduction in sampling cost that itself grows with lattice size. This reduction is negligible at $m \leq 3$, a few percent at $m = 4, 5$ and greatest at $m = 6$, where mitigation with all stabilizers introduces an approximately 27\% increase in cost compared to a partial configuration.

\section{Discussion} \label{sec:disc}
QEM techniques which integrate with QEC and QED at the physical qubit level offer a promising approach to increase computation accuracy on noisy hardware.
This is supported by a growing number of studies which establish that the sampling cost of QEM can be decreased by relying on QEC or QED to suppress a subset of circuit errors.

Our GHZ state simulations show that the benefit of \qedpec is determined by both stabilizer choice and circuit structure.
Our classical optimization over stabilizer measurement configurations shows that, for linear-depth preparation circuits, no sequence of $Z_i Z_{i+1}$ GHZ stabilizer generators enables \qedpec to outperform PEC in sampling cost when QEM is also applied to remove stabilizer measurement bias.
Instead, we find that \qedpec requires non-local stabilizers, such as $Z_0 Z_{n-1}$, to exhibit an advantage, 
with $\textrm{TSE}$ reduced by roughly an order of magnitude compared to PEC.
Such stabilizer circuits coincide with a known fault-tolerant GHZ preparation circuit (e.g.\ Fig.~5(a) in~\cite{gotoManyhypercubeCodesHighrate2024}) and the need for non-locality connects to structural features identified in quantum LDPC codes~\cite{berthusenPartialSyndromeMeasurement2024}.

Repeating the experiments with logarithmic-depth preparation circuits reveals a further connection between circuit structure and the performance of \qedpec. 
Because these shallower circuits are less affected by idling noise, any stabilizer configuration contributes a proportionally larger share of the total noise. 
Moreover, error propagation through these circuits produces a more uniform distribution of detectable errors across the output qubits than in the linear-depth case (see Fig.~\ref{fig:ghz_error_prop}). 
Consequently, we observe a much more restricted advantage for \qedpec over PEC in this setting.

Detecting most errors with the relatively small number of low-weight stabilizers that remain feasible at the error probabilities considered in our simulations is therefore easier in the linear-depth case, where detectable errors are concentrated on a small subset of qubits. 
These findings suggest that QEM should be tailored to circuit structure: circuits in which detectable errors are localised to a few qubits may better justify the noise overhead of stabilizer measurements than circuits in which such errors are more broadly distributed.

Simulations of the time dynamics of the $2 \times 2$ spinless FH model show a more robust advantage for \qedpec , which gives higher accuracy for a range of lattice operator observables (Fig.~\ref{fig:fh_res_main_error}) and for the two-point charge correlator (Fig.~\ref{fig:norm_cc}).
The advantage over PEC increases with evolution time and therefore with circuit depth. 
In contrast, for the radial charge correlator (Fig.~\ref{fig:radial_cc}), whose values lie much closer to zero than those of the two-point correlator, \qedpec and PEC generally perform comparably.

The advantage of \qedpec arises when all available stabilizing loop operators are measured once at the end of computation.
In our sampling cost scaling analysis, see Fig.~\ref{fig:fh_res_main_cost}, we find that for $2 \leq m \leq 3$, our optimisation continues to suggest measuring all loop operators. 
For $4 \leq m \leq 6$, measuring a subset of loop operators minimizes the \qedpec sampling overhead, but the cost reduction over PEC is small.

Previous studies show that \qedpec is effective for FH model simulation with a range of fermion-to-qubit mappings~\cite{caiMultiexponentialErrorExtrapolation2021, papicNearTermFermionicSimulation2025}. Our results extend this evidence to the GSE.
The fault-tolerant properties of the GSE loop operator measurement circuits, and their shallow depth in comparison to time evolution circuits, suggest that multiple rounds of stabilizer measurements could further improve the accuracy of \qedpec, as observed for other mappings~\cite{papicNearTermFermionicSimulation2025}. 
Future work may thus leverage our optimisation framework to determine the placement of stabilizer measurements that minimises the sampling cost of \qedpec.
Further sampling cost reductions could come from applying PEC only to errors that are undetectable, in the lightcone of given observable and in combination with our symmetry optimization~\cite{kimEvidenceUtilityQuantum2023, eddinsLightconeShadingClassically2024}.

The model of undetectable noise we use omits higher order terms involving the product of $k > 1$ detectable errors at each layer and between noisy layers.
Higher order errors at a single layer can be handled at the cost of additional classical processing, see App.~\ref{app:qedpec_high_ord} for an approach involving $\mathcal{O}(2^k)$ steps.
Reducing this computational cost and handling products of detectable errors across layers represent important directions to minimize \qedpec bias.

Developing techniques to check error detectability for more general circuits may be necessary to use \qedpec more broadly.
We consider Clifford circuits and circuits with structure which enabled efficient error detectability checks.
For circuits where propagation through non-Clifford operations cannot be ignored, but the overall circuit is expected to lie in a known symmetry sector, a possible approach is to approximately calculate $\langle S_E \rangle = \textrm{tr}\{ S U_L \cdots E \cdots U_1 \rho U_1^{\dagger} \cdots E \cdots U_L^{\dagger}\}$ using a Pauli-propagation algorithm \cite{begusicSimulatingQuantumCircuit2023, rudolphPauliPropagationComputational2025}. 
This may suffice if we only require confidence that $\langle S_E \rangle$ is positive or negative rather than precisely $\pm 1$. 

More generally, ensuring that the advantages we identify are preserved under realistic hardware constraints, such as mapping to architectures with restricted connectivity through SWAP networks as in~\cite{haggeErrorMitigationError2023}, will be important for deploying \qedpec in practice.
Systematic approaches to discovering fault-tolerant compilations~\cite{rodatzFaultToleranceConstruction2025} or circuit symmetries~\cite{labordeQuantumAlgorithmsTesting2022} 
in combination with optimisation over symmetry measurements 
could significantly extend the range of circuits where \qedpec is applicable.

\section{Data Availability}

The data that support the findings of this study are available from TO upon reasonable request.

\section{Code Availability}
The code that support the findings of this study are available from TO upon reasonable request.

\section{Acknowledgements}
The authors thank Laurin E. Fischer for useful discussions.
TO acknowledges support by the EPSRC through an EPSRC iCASE studentship award in collaboration with IBM Research (EP/W522211/1).
DJ is partly funded by the Cluster of Excellence `Advanced Imaging of Matter' of the Deutsche Forschungsgemeinschaft (DFG)|EXC 2056- project ID390715994. 
DJ acknowledges support by the DFG project ``Quantencomputing mit neutralen Atomen" (JA 1793/1-1, Japan-JST-DFG-ASPIRE 2024), and the Hamburg Quantum Computing Initiative (HQIC) project EFRE. 
The EFRE project is co-financed by ERDF of the European Union and by ``Fonds of the Hamburg Ministry of Science, Research, Equalities and Districts (BWFGB)."
This work was supported as a part of NCCR SPIN, a National Centre of Competence in Research, funded by the Swiss National Science Foundation (grant number 225153).
The authors would like to acknowledge the use of the University of Oxford Advanced Research Computing (ARC) facility in carrying out this work~\cite{richardsUniversityOxfordAdvanced2015}.

\bibliography{bibliography}

\appendix

\newpage
\onecolumngrid
\section{Additional background on quantum error mitigation techniques}

\subsection{Probabilistic error cancellation} \label{app:pec}

The noisy implementation of an ideal $n$ qubit unitary $U$ can be modelled as the noise free gate followed by a quantum channel: $\Lambda \circ \mathcal{U} (\rho)$ where $\mathcal{U} = U \rho U^{\dagger}$. We assume that $\Lambda$ can be written as a Pauli channel : 
\begin{align}
    \Lambda(\rho) = \sum_{i=1}^{4^n} p_i E_i \rho E_i^{\dagger},
\end{align}
with $\sum_i p_i = 1$ and $p_i \geq 0$. This assumption can be justified in practice by using Pauli twirling \cite{wallmanNoiseTailoringScalable2016}.

We define $\Lambda^{-1}$ as the map which satisfies $\Lambda^{-1} \circ \Lambda (\rho) = \rho$. 
We can write 
\begin{align}
    \Lambda^{-1} (\rho) = \sum_{i=1}^{4^n} c_i E_i \rho E_i^{\dagger},
\end{align}
where we can have $c_i < 0$ and $\sum_i c_i > 1$. In general, $\Lambda^{-1}$ in this form is not completely positive or trace preserving and therefore does not define a physical quantum channel. However, by defining  
\begin{align}
    \gamma = \sum_i \lvert c_i \rvert,\\
    p^{inv}_i = \lvert c_i \rvert / \gamma,
\end{align}
we can write 
\begin{align}
    \Lambda^{-1} (\rho) = \gamma \sum_i {\rm sgn} (c_i) p^{inv}_i E_i \rho E_i,
\end{align}
 where ${\rm sgn}$ denotes the sign function, $\sum_i p^{inv}_i = 1$ and $p^{inv}_i \geq 0$. 
PEC then enables unbiased estimation of expectation values following the same procedure described in Sec.~\ref{sec:background} in the man text. 
A distinction to applying PEC to the Pauli-Lindblad model is that Pauli strings are sampled from the inverse channel $E_i \sim p^{inv}_i$ instead of combining multiple Bernoulli draws.

The mean squared error (MSE) captures both the bias and variance increase of an error mitigated estimator
\begin{align}
    \text{MSE}[\overline{O}_{\scriptscriptstyle \text{PEC}}] = \text{Bias}[\overline{O}_{\scriptscriptstyle \text{PEC}}]^2 + \text{Var}[\overline{O}_{\scriptscriptstyle \text{PEC}}].
\end{align}
Because $\text{Bias}[\overline{O}_{\scriptscriptstyle \text{PEC}}] = \mathbb{E}[\overline{O}_{\scriptscriptstyle \text{PEC}}] - \langle O_{\text{ideal}} \rangle = 0$, the MSE reduces to the variance
\begin{align}
    \text{MSE}[\overline{O}_{\scriptscriptstyle \text{PEC}}] = \text{Var}[\overline{O}_{\scriptscriptstyle \text{PEC}}],
\end{align}
which is scaled by $C_{\scriptscriptstyle\text{PEC}}$ compared to the unmitigated sample mean $\overline{O}_{\text{noisy}}$
\begin{align}
    \text{MSE}[\overline{O}_{\scriptscriptstyle \text{PEC}}] = C_{\scriptscriptstyle \text{PEC}} \text{Var}[\overline{O}_{\text{noisy}}].
\end{align}

We restrict our treatment of PEC to constructing an inverse channel using Pauli channels. In general one considers the problem of synthesizing noise free quantum gates from a noisy basis available on device. That is, how to choose the coefficients $\{ c_i \}$ such that $\mathcal{U} = \sum_i c_i \mathcal{B}_i$ where $\mathcal{U}$ is the target noise-free gate and $\mathcal{B}_i \in \{ \Lambda_i \circ \mathcal{U}_i \}$. The set of coefficients with minimal 1-norm can be found using semi-definite programming \cite{temmeErrorMitigationShortdepth2017}. With a Pauli inversion basis, we consider a basis consisting of channels $\Lambda \circ \mathcal{P}_j \circ \mathcal{U}_i$ for $0 \leq j \leq 4^{n - 1}$. Recovery noise introduced by $\mathcal{P}_j$ could be neglected if it is small in comparison to $\Lambda$. Alternatively, if $\Lambda$ is a Pauli channel, propagating it through  $\mathcal{P}_j$ to combine with the recovery Pauli noise channel and adjusting the quasi-probability decomposition accordingly. In the following we assume noise following PEC recovery operations is sufficiently small to neglect.

\subsection{Quantum error detection} \label{app:qed}
In the simple case of detection being performed immediately after an $n$ qubit Pauli noise channel 
\begin{align}
    \Lambda (\rho) = \sum_{i = 0}^{4^n - 1} p_i E_i \rho E_i,
\end{align}
the likelihood that no error is detected is 
\begin{align}
    p_{\nd} &= p_0 + \sum_{i : E_i \in \mathcal{E}_{\text{undet}}} p_i\\
            &= 1 - p_{\detec},
\end{align}
where $p_0$ is the probability of no error occurring and $p_{\detec}$ is the probability of an error being detected, depending on the set of symmetries measured. 
For circuits with detection performed after multiple layers, it is necessary to consider the probability of errors which remain detectable after propagation through the intermediate operations. 
This propagation can be performed classically for Clifford circuits and 
certain Clifford~+~$R_P(\theta)$ circuits, as described in more detail in Sec.~\ref{sec:qedpec}, or circuits designed fault-tolerantly. 
For a depth $L$ circuit where the probability of a detectable error occurring at each layer is uniformly $p_{\detec}$, if the probability that an error at a given layer remains detectable at the point of error detection is $f$, then the probability an error will be detected after $L$ layers is $p_{\detec}^{\text{final}} = 1 - (1 - f p_{\detec})^L \approx 1 - e^{- L f p_{\detec}}$. Then $p_{\nd}^{\text{final}} = 1 - p_{\detec}^{\text{final}}$ and the QED sampling cost can be approximated as
\begin{align} \label{eq:qed_cost_tot}
C_{\text{QED}, L} \approx e^{L f p_{\detec}}.
\end{align}
QED is the most sample efficient known QEM technique, alongside the best-case performance of Verified Phase Estimation or Echo Verification \cite{obrienErrorMitigationVerified2021, obrienPurificationbasedQuantumError2023} or rescaling under global depolarizing noise \cite{tsubouchiSymmetricCliffordTwirling2025}.

Writing the MSE of the QED estimator, we obtain
\begin{align}
    \text{MSE}[\overline{O}_{\scriptscriptstyle \text{QED}}] = \text{Bias}[\overline{O}_{\scriptscriptstyle \text{QED}}]^2 +  C_{\scriptscriptstyle \text{QED}} \, \text{Var}[\overline{O}_{\text{noisy}}],
\end{align}
where $\text{Bias}[\overline{O}_{\scriptscriptstyle \text{QED}}] = \mathbb{E}[\overline{O}_{\scriptscriptstyle \text{QED}}] - \langle O_{\text{ideal}} \rangle$ and $\text{Var}[\overline{O}_{\text{noisy}}]$ denotes the variance of a standard sample-mean estimator at fixed total number of circuit executions.
This expression makes explicit the trade-off underlying QED as an error-mitigation method: detectable errors can be suppressed at the cost of increased sampling overhead due to post-selection.

It has been shown that QED can be implemented virtually by reconstructing a projector onto the desired symmetry subspace instead of measuring all symmetries and discarding \cite{caiQuantumErrorMitigation2021, tsubouchiVirtualQuantumError2023}. This approach has a quadratically higher sampling cost than post-selection QED but may reduce the circuit depth and qubit requirements for implementation.

\subsection{Efficient error propagation \label{app:error_prop}}
In general, propagation through a non-Clifford gate $U$ converts a Pauli error $E$ into a general unitary $E U = U (U^{\dagger} E U)$, where the propagated unitary can be decomposed into Pauli operators $(U^{\dagger} E U) = \sum_i c_iP_i$, where $c_i$ depends on $U$ and $E$. Repeated propagation through multiple circuit layers leads to a Pauli decomposition which grows exponentially fast with circuit size. 
Therefore, the detectability of Pauli errors at a given circuit layer cannot generally be checked efficiently by propagating through circuits with many non-Clifford layers. 
However, for circuits where reversing the order of a Pauli error $E$ and a non-Clifford layer $U$ gives $U E = E' U'$, for a Pauli error $E'$ and a possibly modified non-Clifford layer $U'$, propagated error detectability can be efficiently checked if $[U', S_i] = 0 \ \forall \ i$. 
This is the case for non-Clifford layers with gates of the form $R_P(\theta) = \exp(-\mathrm{i} \theta P)$ for some Pauli $P$ if in the absence of noise the circuit remains in the same symmetry sector for any $\theta$ \cite{papicNearTermFermionicSimulation2025}. 
This is because propagating an error $E$ through $R_P(\theta)$ does not change $E$
\begin{align}
    R_P(\theta) E = E R_P((-1)^{\langle E, P \rangle}\theta),
\end{align}
where $\langle E, P \rangle = 1$ if $E$ and $P$ commute, else -1. 
Because the circuit remains in the same symmetry sector for any $\theta$, the outcome of error-detection will only be affected by $E$. Therefore, it suffices to only propagate errors through Clifford gates. 

\subsection{Symmetry informed stabilizer circuit mitigation \label{app:sym_circ_mit}}
\begin{figure}
    \centering
    \includegraphics[width=0.75\linewidth, trim={0mm 8mm 0 5mm}, clip]{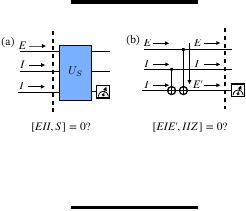}
    \caption{Different approaches to check whether a Pauli error $E$ on the upper wire can be detected after circuit propagation. (a) check whether $E$ commutes with Pauli symmetry $S$ which is measured using the circuit $U_{S}$ and $I$ denotes the identity. (b) Decompose $U_{S}$ and check whether the error will commute with the expected deterministic measurement. This enables a more accurate undetectable noise channel and enables symmetry-informed mitigation of noise in $U_{S}$.}
    \label{fig:sym_meas}
\end{figure}

In addition to optimizing the stabilizer measurement configuration, the structure of stabilizer measurement circuits $U_{\scriptscriptstyle \textrm{SYM}}$ can be taken into account when checking error detectability.
By performing additional classical error propagation this enables more accurate construction of $\Lambda_{\text{undet}}$ and enables application of symmetry-informed PEC to symmetry measurement circuits themselves.
In Fig.~\ref{fig:sym_meas}(a) $U_{\scriptscriptstyle \textrm{SYM}}$ is treated as a black box. $E$ is undetectable if $[E, S] = 0$ for all symmetries $S$. Noise in $U_{\textrm{SYM}}$ is mitigated using PEC.
In Fig.~\ref{fig:sym_meas}(b) $U_{\scriptscriptstyle \textrm{SYM}}$, in this case the measurement circuit of $\langle ZZ \rangle$ is decomposed into 2 CX gates. Instead of viewing the expectation $\langle ZZ\rangle$ as the invariant quantity, the ancilla measurement outcome is viewed as invariant. Errors are defined as detectable if they anti-commute with this measurement after propagation. Errors within $U_{\scriptscriptstyle \textrm{SYM}}$ can then be defined as detectable or undetectable. Noise in $U_{\scriptscriptstyle \textrm{SYM}}$ is now mitigated using \qedpec.

\subsection{Higher order cancellation in \qedpec} \label{app:qedpec_high_ord}
In the presentation of \qedpec in Sec.~\ref{sec:qedpec} errors are only being cancelled to first order.
This is because residual undetectable errors introduced by 2 or more detectable errors occurring will introduce bias into the mitigated estimator. 
Higher order undetectable errors are missed at: 1. the conversion of $\Lambda$ to $\Lambda_{\text{undet}}$ for the Sparse Pauli Lindblad model; 2. when multiple errors occur in the circuit, and after propagation to the point of symmetry measurement their product is undetectable.
The bias introduced through 1. can be reduced by expanding out terms in the Pauli-Lindblad noise model to include 2nd or higher order undetectable errors, then constructing an inverse channel to also target these higher order terms. 
The sample cost of inverting undetectable errors to order $\kappa$ can be approximated to within $(1 - w)^{\kappa+1}$ error. This is done by expanding out noise model terms to include products of $\kappa$ errors, then approximately factoring this expansion back into a composition of binary outcome channels where each error occurs with the same probability as in the expanded channel. 
Exact identification of all orders of undetectable errors would require full noise model expansion, which has $2^\mathcal{K}$ terms. This is not computationally scalable, therefore methods to identify higher order undetectable errors and applications where these matter is a valuable direction for future research.

\section{Probability of undetectable errors for
single uniform weight Pauli-Lindblad noise model}
 \label{app:deriv_undet}
The Sparse-Pauli-Lindblad (SPL) noise model $\Lambda$ with a generating set $\mathcal{K}$ can be written 
\begin{align} \label{eq_app:spl}
    \Lambda (\rho) = \bigcirc_{k \in \mathcal{K}} \left( w_k \cdot + (1 - w_k) E_k \cdot E_k \right) (\rho),
\end{align}
where $1 - w_k$ is the probability of the Pauli error $E_k$ occurring.
We denote the set of possible errors that can be generated by this channel as $E_{\mathcal{K}}$, where
\begin{align}
    E_{\mathcal{K}} = \left\{E(b) = \prod_{j=1}^{\lvert \mathcal{K} \rvert}E^{b_j}_k \ \lvert \ b \in \{0, 1\}^{\lvert \mathcal{K} \rvert} \right\}.
\end{align}
For an $n$-qubit circuit with a set of symmetries that can be specified by a stabilizer group $\mathcal{S}$, where $\mathcal{S}$ is an Abelian subgroup of the Pauli group $\mathcal{P}_n$ with $-I \notin \mathcal{S}$, we define the centralizer of $\mathcal{S}$ in $\mathcal{P}_n$ as 
\begin{align}
    \mathcal{C}(\mathcal{S}) = \{E\in \mathcal{P}_n \lvert \ \forall \ S \in \mathcal{S},\ SE = ES \},
\end{align}
as $\mathcal{C}(\mathcal{S})$ contains $\mathcal{S}$, we take the set of undetectable errors $\mathcal{E}_{\text{undet}}$ to be 
\begin{align}
    \mathcal{E}_{\text{undet}} = \mathcal{C}(\mathcal{S}) \ \backslash 
    \ \mathcal{S}.
\end{align}
The set of detectable errors $\mathcal{E}_{\text{det}}$ is then 
\begin{align}
    \mathcal{E}_{\text{det}} = \{E \in \mathcal{P}_n \ \vert \ E \notin  \mathcal{C}(\mathcal{S})\}.
\end{align}

We then define the set of undetectable error generators as 
\begin{align}
   \mathcal{K}^{\text{undet}} = \{E_k \ \vert \ k \in \mathcal{K},\ E_k \in \mathcal{E}_{\text{undet}} \},
\end{align}
and the set of undetectable error generators as 
\begin{align}
   \mathcal{K}^{\text{det}} = \{E_k \ \vert \ k \in \mathcal{K}, \ E_k \in \mathcal{E}_{\text{det}} \}.
\end{align}

\textit{Single symmetry.}
To calculate the probability that no error is detected after $\Lambda (\rho)$ acts and we measure a single symmetry, we calculate the probability that 0 errors, 1 or more undetectable errors and 2 or more even numbers of detectable errors occur.
Because the channel is constructed as a composition of independent sub-channels, we note that undetectable and detectable error generators occur independently. 
The probability that 0 or more undetectable errors occur is 1. 
Therefore we need to calculate only the probability that 0 or more even numbers of detectable errors occur. 
For an SPL model with uniform weight $\lambda$, there is a binomial distribution over outcomes. 
If we define $L = \lvert \mathcal{K}^{\text{det}} \rvert$, we therefore want the probability of an even number of positive binomial draws $p_{\text{even}}$, where each draw is positive with probability $p = 1 - w = (1 + e^{- 2 \lambda}) / 2$.
This probability is then 
\begin{align}
    p_{\text{even}} &= \frac{(p + (1 - p))^L + ((1 - p) - p)^L}{2}\\
    &= \frac{1 + (1 - 2p)^L}{2}\\
    &= \frac{1 + (2w - 1)^L}{2},
\end{align}
where we have obtained a even-term binomial expansion by taking the sum of alternating sign expansions $(a + b)^L + (a - b)^L$. 
Because $2w - 1 = e^{- 2 \lambda}$, we then have 
\begin{align}
  p_{\text{even}} &= \frac{1 + e^{- 2 \lambda L}}{2}.
\end{align}

\textit{Multiple Symmetries}.
The probability of no detectable error occurring when measuring multiple symmetries is the total likelihood an error $E \in \mathcal{E}_{\text{undet}}$ occurs after $\Lambda (\rho)$
\begin{align}
    p_{\nd} = \sum_{\omega = 0}^{\lvert \mathcal{K} \rvert} w^{\lvert \mathcal{K} \rvert - \omega} (1 - w)^{\omega} \mathcal{W}_{\omega},
\end{align}
where $\mathcal{W}_{\omega}$ is the weight-enumerator
\begin{align}
    \mathcal{W}_{\omega} = \lvert \{ E(b) \in E_{\mathcal{K}} \cap \mathcal{E}_{\text{undet}} \ \vert \ \text{wt}(b) = \omega \} \rvert,
\end{align}
which counts the number of errors generated by $\Lambda$ involving a product of $\text{wt}(b)$ non identity Pauli operators where $b$ is the binary string specifying elements of $E_{\mathcal{K}}$.  
The exact form of $p_{\nd}$ will depend on the weight enumerator which in turn depends on the symmetries present and error model.  
To first order in error probability $(1 - w)$ we have
\begin{align}
    p_{\nd} &= 1 - (\lvert \mathcal{K} \rvert - \lvert \mathcal{K}_{\text{undet}} \rvert)(1 - w) + \mathcal{O}((1 - w)^2)\\
    &\approx 1 - \lvert \mathcal{K}_{\text{det}} \rvert (1 - w)\\
    &= 1 - \frac{\lvert \mathcal{K}_{\text{det}} \rvert}{2}(1 - e^{-2 \lambda}).
\end{align}
Alternatively we can approximate the probability of no error being detected as the probability that no detectable error occurs.
This may be slightly more accurate than the above truncation based approximation as it accounts for products of undetectable errors occurring and ignores 2nd order and higher products of detectable errors.
In this case
\begin{align}
    p_{\nd} &= w^{\lvert \mathcal{K}_{\text{det}} \rvert}\\
    &= \left(\frac{1 + e^{-2 \lambda}}{2}\right)^{\lvert \mathcal{K}_{\text{det}} \rvert}.
\end{align}

\section{Simulation details} \label{app:sim}

\subsection{Noise model}

The superconducting inspired noise model with error rate $p$ applies 1-qubit depolarising noise after each 1-qubit gate with probability $p/10$ and 2-qubit depolarising noise after each 2-qubit gate with probability $p$. The ratios of different gate errors are derived from a superconducting-inspired noise model and chosen to reflect error rates of current hardware.
Noise is not added to $R_Z(\theta)$ gates as they can be implemented in software using frame changes~\cite{mckayEfficientGatesQuantum2017}.
In terms of Pauli-Lindblad model generators $\mathcal{K}$, this corresponds to $\{X, Y, Z\}$ for one-qubit gate noise and $\{IX, IY, IZ, XI, YI, ZI, XX, XY, XZ, YX, YY, YZ, ZX, ZY, ZZ \}$ for two-qubit gate noise.
Each error probability $p_k$ is converted to a Pauli-Lindblad model weight $\lambda_k$ using $p_k = (1 - \exp(-2 \lambda_k)) / 2$.
Prior to measurement, bitflip errors occur with probability $p$.
Qubit resets are followed by 1-qubit depolarising noise of probability $2p$.

\subsection{Gates}
The basis gate set consists of arbitrary single qubit rotations, CX gates, arbitrary Pauli-controlled-Pauli gates, qubit reset and measurement.

\section{GHZ state preparation results without idling noise \label{app_ghz}}
Here we include the results of our numerical experiments for GHZ state preparation with the same setup and noise model as in the main text, but without idling noise.
During stabilizer selection for linear-depth circuits, we find the same structure is identified as with idling. We reproduce these results here
\begin{align} \label{eq:app_ghz_stab}
    I_0 Z_1 I_2 \cdots I_{n-3} Z_{n-2} I_{n-1},\\
    Z_0 I_1 I_2 \cdots I_{n-3} I_{n-2} Z_{n-1},\\ 
    Z_0 I_1 I_2 \cdots I_{n-3} Z_{n-2} I_{n-1},\\
    \label{eq:app_ghz_stab_end}
    I_0 Z_1 I_2 \cdots I_{n-3}I_{n-2} Z_{n-1},
\end{align}
with a single stabilizer selected out of these degenerate solutions. The optimization scores for these solutions are in Tab.~\ref{tab:app_ghz_lin_scores}.

In stabilizer selection for logarithmic depth circuits, we observe a distinct set of stabilizers is selected compared to experiments with idling noise, as shown in Tab.~\ref{tab:app_ghz}, which also includes optimization scores.

\begin{table}[]
    \centering
    \begin{tabular}{c | c | c | c}
      $n$  &  $ \left(\gamma_{\scriptscriptstyle \text{undet}} \right)^2 \times $ & $\left(\gamma_{\scriptscriptstyle \text{undet}} \right)^2 \times $  & $\gamma^2$ \\ 
       & $\ \ C_{\scriptscriptstyle \textrm{SYM}}(\mathcal{C})$ & $C_{\scriptscriptstyle \textrm{SYM}}(\mathbb{G}_{\scriptscriptstyle\textrm{GHZ}})$ & 
      \\\hline 
     7   & 1.02 & 1.11 & 1.02 \\
     10  & 1.03 & 1.17 & 1.04 \\
     15  & 1.04 & 1.27 & 1.06 \\
     20  & 1.05 & 1.38 & 1.08 \\
     30  & 1.07 & 1.62 & 1.12 \\
     40  & 1.09 & 1.90 & 1.17 \\
     50  & 1.11 & 2.24 & 1.22
    \end{tabular}
    \caption{Optimization scores for stabilizers $\mathcal{C}$ in Eqs.~(\ref{eq:app_ghz_stab}-\ref{eq:app_ghz_stab_end}) for procedure  in Sec.~\ref{sec:sym_selec_a} in the main text applied to a GHZ state preparation circuit with depth growing linearly in size $n$ (excluding terminating computational basis measurements). 
    Optimization scores are given by the sampling cost of applying PEC to undetectable errors $\left(\gamma_{\scriptscriptstyle \text{undet}} \right)^2 $ and stabilizer measurement circuit errors $C_{\scriptscriptstyle \textrm{SYM}}(\mathcal{C})$.
    When $\mathcal{C} = \mathbb{G}_{\scriptscriptstyle\textrm{GHZ}}$ all stabilizer generators are used. $\gamma^2$ is the cost of applying PEC to all state preparation errors.
    }
    \label{tab:app_ghz_lin_scores}
\end{table}

\begin{table}[]
    \begin{tabular}{c | c | c | c | c}
      $n$  & $\mathcal{C}$ & $ \left(\gamma_{\scriptscriptstyle \text{undet}} \right)^2 \times $ & $\left(\gamma_{\scriptscriptstyle \text{undet}} \right)^2 \times $  & $\gamma^2$ \\ 
       &  & $\ \ C_{\scriptscriptstyle \textrm{SYM}}(\mathcal{C})$ & $C_{\scriptscriptstyle \textrm{SYM}}(\mathbb{G}_{\scriptscriptstyle\textrm{GHZ}})$ & 
      \\\hline 
     15 & $Z_0 Z_{11}$ & 1.05 & 1.27 & 1.06 \\
     20 &  $Z_0Z_{19}$, $Z_{5}Z_{14}$ & 1.07 & 1.38 & 1.08 \\
     30  & $Z_0Z_{19}$, $Z_{10}Z_{26}$ & 1.11 & 1.61 & 1.12 \\
     40 & $Z_6Z_{20}$, $Z_{8}Z_{35}$, $Z_{14}Z_{26}$ & 1.15 & 1.90 & 1.17 \\
     50 & $Z_2Z_{27}$, $Z_{14}Z_{32}$, $Z_{16}Z_{43}$ & 1.19 & 2.24 & 1.22
    \end{tabular}
    \caption{Stabilizer configurations $\mathcal{C}$ selected by the optimization in Sec.~\ref{sec:sym_selec_a} in the main text for a GHZ state preparation circuit with depth growing logarithmically in size $n$ (excluding terminating computational basis measurements) without idling noise.
    Optimization scores are given by the sampling cost of applying PEC to undetectable errors $\left(\gamma_{\scriptscriptstyle \text{undet}} \right)^2 $ and stabilizer measurement circuit errors $C_{\scriptscriptstyle \textrm{SYM}}(\mathcal{C})$.
    When $\mathcal{C} = \mathbb{G}_{\scriptscriptstyle\textrm{GHZ}}$ all stabilizer generators are used. $\gamma^2$ is the cost of applying PEC to all state preparation errors.
    }
    \label{tab:app_ghz}
\end{table}

\begin{figure*}[ht]
  \centering
  \begin{tabular}{@{}ccc@{}}
    \includegraphics[width=0.32\textwidth]{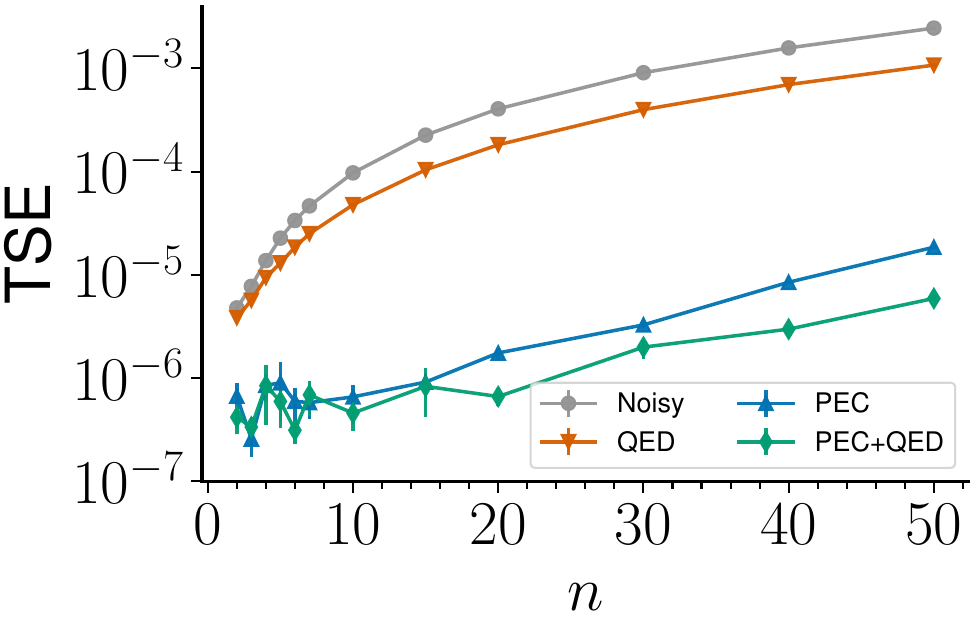} 
    \begin{picture}(0,0)
        \put(-160,108){\scriptsize (a)} 
    \end{picture}
    
    &
    \includegraphics[width=0.32\textwidth]{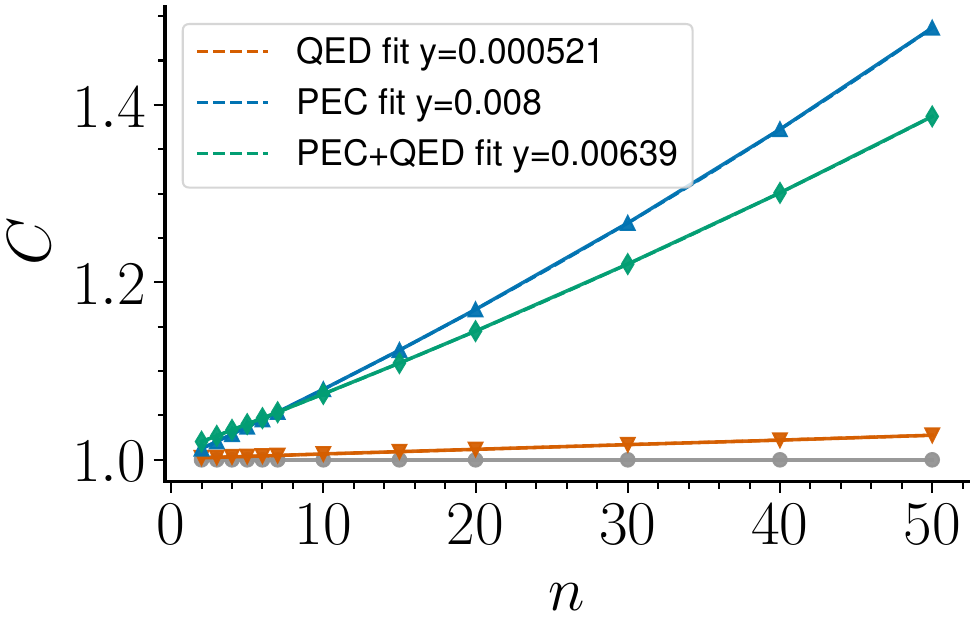} 
    \begin{picture}(0,0)
        \put(-160,108){\scriptsize (b)} 
    \end{picture}
    
    &
    \includegraphics[width=0.32\textwidth]{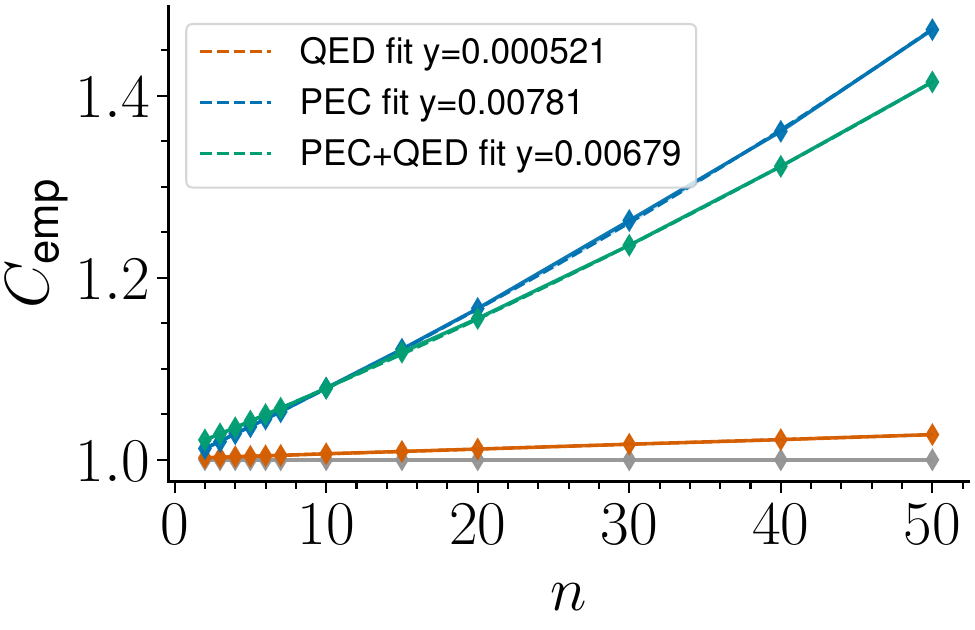}
    \begin{picture}(0,0)
        \put(-160,108){\scriptsize (c)} 
    \end{picture}
    
    \\

     \includegraphics[width=0.32\textwidth]{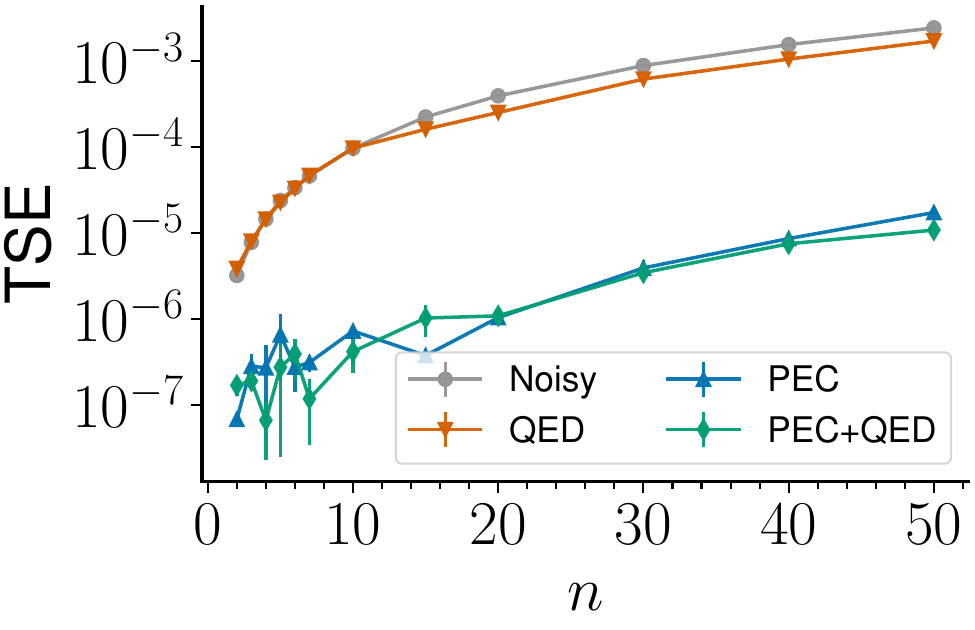} 
    \begin{picture}(0,0)
        \put(-160,108){\scriptsize (d)} 
    \end{picture}
    
    &
    \includegraphics[width=0.32\textwidth]{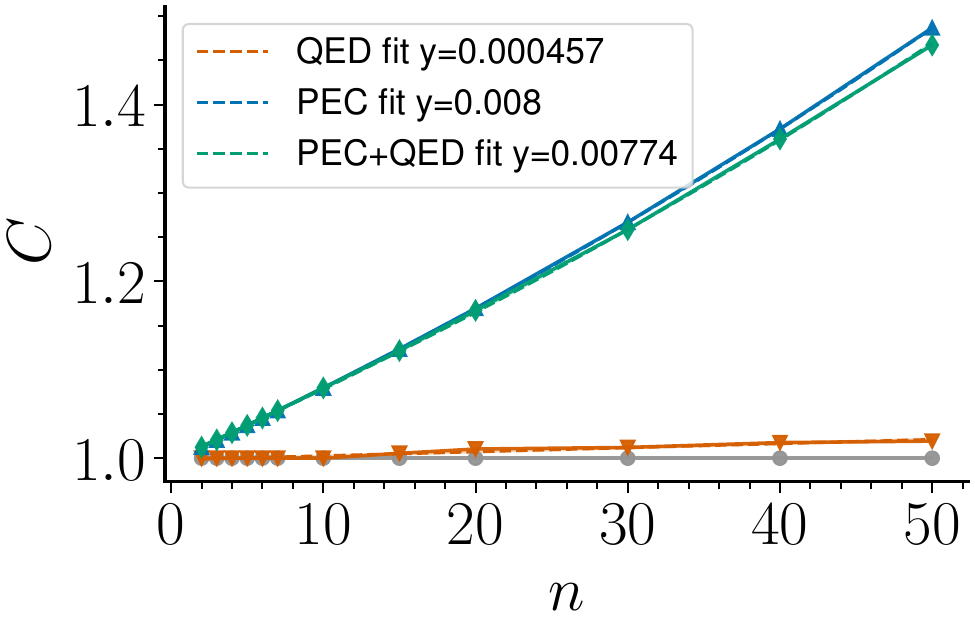} 
    \begin{picture}(0,0)
        \put(-160,108){\scriptsize (e)} 
    \end{picture}
    
    &
    \includegraphics[width=0.32\textwidth]{figures/ghz_cost_scaling_pred_no_idle_log1_propii_errorbars.pdf}
    \begin{picture}(0,0)
        \put(-160,108){\scriptsize (f)} 
    \end{picture}
  \end{tabular}
  \caption{GHZ state preparation mitigation for linear and logarithmic depth circuits.
  The total square error (TSE) and mitigation costs between the output of a noise-free and noisy GHZ state preparation circuit with increasing GHZ state size $n$. 
  Results obtained using stabilizer circuit simulations with: no mitigation ``Noisy" (grey), QED (orange), PEC (blue) and \qedpec (green), all simulations use $10^6$ shots.
  Results are shown for linear  (a), (b), (c)  and logarithmic depth (d), (e), (f) circuits.
  Mitigation costs are fit to an exponential $C(n) = x e^{y n} + z$ with $x + z \sim 1$.
  $C_{\textrm{pred}}$ is the predicted sampling cost using PEC theory and empirical QED discard rate.
  $C_{\textrm{emp}}$ is the empirical sampling cost obtained by calculating the total number of discarded shots after applying PEC to experiment counts.}
  \label{fig:ghz_res_no_idle}
\end{figure*}

In Fig.~\ref{fig:ghz_res_no_idle}  we see that for the linear-depth circuit, without idling noise, at small $n < 10$ the sampling overhead advantage of PEC displayed in (b) and (c) is too small to be resolved, as both PEC and \qedpec have reached a finite-shot noise floor. 
At larger $n$, the smaller sampling overhead of \qedpec appears as just under an order of magnitude reduction in TSE compared to PEC. 
In Fig.~\ref{fig:ghz_res_no_idle}(d), we see a small advantage for \qedpec can obtained without idling noise at $n > 20$.
We find the predicted sampling cost in (e) is slightly smaller than the empirically observed sampling cost in (f). 

\section{Example of a generalized superfast encoded Hamiltonian} \label{app:gse_2x2}
\begin{figure}
    \centering
    \includegraphics[width=0.49\textwidth, trim=0 3 0 0, clip]{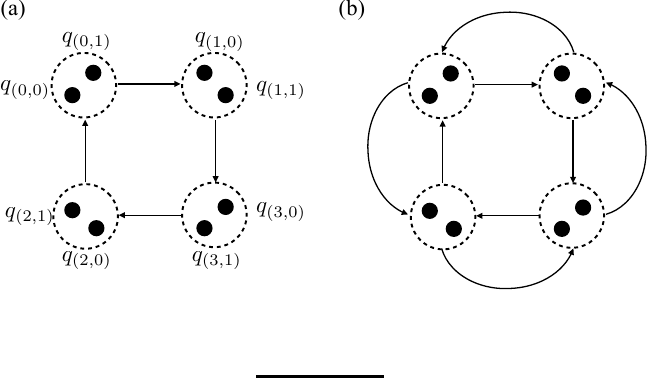}
    \caption{(a) Qubit indexing for $2 \times 2$ lattice. (b) Addition of boundary loops to ensure nodes have degree 4.}
    \label{fig:index}
\end{figure}

As a guiding example, we will construct the qubit Hamiltonian for the $2\times2$ spinless model by applying the definitions from \cite{haggeErrorMitigationError2023}.
The first step of this encoding is to decide on generalized Majorana operators to place at each fermion site. 
As these operators are tensor products of Pauli matrices, the only formal requirement is that the set of Majorana operators at each site mutually anti-commute. However, for the resulting system to have error detecting properties there is an additional requirement of the logical operators defined by the Majorana operators having a minimum weight of 2. 
One may try and place 2 qubits at each vertex of the fermion interaction graph, as shown in Fig.~(\ref{fig:index})(a), and assign a Majorana operator to each outgoing edge.
Labelling the $p$th Majorana operator leaving the $j$th vertex as $\gamma_{(j, p)}$, we define the logical vertex operator as $B_j = (-\mathrm{i})^{d(j) / 2} \prod_{p=1}^{d(j)} \gamma_{(j, p)}$.
For $B_j$ to have weight 2 and be non trivial, we require that our two Majorana operators have differing Paulis operators in each column, however, this conflicts with our requirement that the Majorana operators anti-commute. Therefore we must consider additional degrees of freedom. 
By adding 2 dummy edges to each vertex in the fermion interaction graph we can choose Majorana operators with a minimum weight logical operator of 2. 
These are dummy edges in the sense that we ignore their contribution to the encoded Hamiltonian, only using them to perform stabilizer measurements. 
This change to the fermion interaction graph can be visualized in Fig.~(\ref{fig:index})(b).
An assignment of Majorana operators at each vertex with a minimum weight 2 logical operator are $\{ \gamma_{(j, 0)}=IZ, \gamma_{(j, 1)}=YY, \gamma_{(j, 2)}=XY, \gamma_{(j, 3)}=IX \}$.

For the $j$th vertex we have a two-qubit vertex operator $B_j = - Z_{(j, 0)} Y_{(j, 1)}$ where $(j, p)$ indexes the $p$th qubit at vertex $j$. 
The edge operators $A_{jk} = \epsilon_{jk} \gamma_{(j, p_{j,k})} \gamma_{(k, q_{k, j})}$ where $\epsilon_{jk} = \pm 1$ is the edge orientation such that $\epsilon_{kj} = - \epsilon_{jk}$ and $(p_{j,k}, q_{k, j})$ index the Majorana operators at the endpoints of the edge $(j, k)$.
As the two Majoranas used to construct each edge operator act on distinct qubits, and each Majorana is of at least one, edge operators always have weight at least two. 
Each edge operator $A_{jk}$ features in the Hamiltonian as $B_j A_{jk}$ and $A_{jk} B_k$. The 8 possible operators of this form are listed in \tab{\ref{tab:vertex_edge_prod}}.
The five distinct loops in the augmented graph correspond to system symmetries we can use for error detection.
The stabilizers are defined using loop operators which, for a loop $\xi$ of length $l$ are defined as $A(\xi) = \mathrm{i}^l \prod_{j=0}^l A_{\xi(j), \xi(j+1)}$ where $\xi(j)$ is a vertex in the loop and $\xi(l) = \xi(0)$.
In our example the four vertex central loop operator is $A(\xi_{(0, 1, 2, 3, 0)}) = I_{(0, 0)} Y_{(0, 1)} X_{(1, 0)} Z_{(1, 1)} Z_{(2, 0)} I_{(2, 1)} Y_{(3, 0)} X_{(3, 1)}$. 

\begin{table}[!h]
   \centering 
    \begin{tabular}{c|c|c} 
         Edge $(j, k)$ & $B_j A_{jk}$ &  $A_{jk} B_k$ \\ \hline
         (0, 1) & $-\mathrm{i} Z_{(0,0)} X_{(0,1)} X_{(1,0)} Y_{(1,1)}$ & 
         $\mathrm{i} I_{(0,0)} Z_{(0,1)} Y_{(1,0)} I_{(1,1)}$ \\
         (1, 3) & 
         $\mathrm{i} Z_{(1,0)} Z_{(1,1)} Y_{(3,0)} Y_{(3,1)}$ & 
         $-\mathrm{i} I_{(1,0)} X_{(1,1)} X_{(3,0)} I_{(3,1)}$ \\
         (3, 2) & $-\mathrm{i} I_{(2,0)} Z_{(2,1)} Y_{(3,0)} I_{(3,1)}$ & 
         $-\mathrm{i} Z_{(2,0)} X_{(2,1)} X_{(3,0)} Y_{(3,1)}$\\
         (2, 0) & $\mathrm{i} I_{(0,0)} X_{(0,1)} X_{(2,0)} I_{(2,1)} $ & 
         $-\mathrm{i} Z_{(0,0)} Z_{(0,1)} Y_{(2,0)} Y_{(2,1)}$ \\ 
    \end{tabular}
    \caption{Products of vertex and edge operators for each edge in the generalized superfast encoded Fermi Hubbard model on a $2 \times 2$ lattice.}
    \label{tab:vertex_edge_prod}
\end{table}

\begin{figure}
    \centering
    \begin{subfigure}(a)    \begin{quantikz}
      \qw &  \gate{P_1} \vqw{3} & \qw & \qw & \qw & \qw &  \qw & \gate{P_1} \vqw{3}  & \qw \\
      \qw & \qw & \gate{P_2}\vqw{2} & \qw & \qw & \qw & \gate{P_2} \vqw{2} & \qw & \qw \\
      \qw & \qw & \qw & \gate{P_3} \vqw{1}& \qw & \gate{P_3} \vqw{1}& \qw & \qw & \qw \\
      \qw & \gate{Q_4} &\gate{Q_4} &\gate{Q_4} & \gate{e^{-\mathrm{i} P_4 t}} & \gate{Q_4} & \gate{Q_4} & \gate{Q_4} & \qw \\
    \end{quantikz}
    \end{subfigure}
    \\[1.5em]
    \begin{subfigure}(b)
    \begin{quantikz}
      \qw &  \gate{P_1} \vqw{3} & \qw & \qw & \qw & \qw &  \qw & \gate{P_1} \vqw{3} & \qw\\
      \qw & \qw & \gate{P_2}\vqw{2} & \qw & \qw & \qw & \gate{P_2} \vqw{2} & \qw & \qw \\
      \qw & \qw & \qw & \qw & \gate[wires=2]{e^{-\mathrm{i} P_3 P_4 t}}  & \qw & \qw & \qw & \qw \\
      \qw & \gate{Q_4} &\gate{Q_4} & \qw & \qw & \qw & \gate{Q_4} & \gate{Q_4} & \qw \\
    \end{quantikz}
    \end{subfigure}
    \caption{Weight-4 Pauli operator rotation circuits.
    Reproduced from Fig.~(11) in \cite{haggeErrorMitigationError2023}. 
    Circuits for evolution of $\exp(-\mathrm{i}Pt)$ 
    where $P = P_1 P_2 P_3 P_4$, $P_i$ are Pauli matrices and
    $Q_4$ is a Pauli matrix such that $[P_4, Q_4] \neq 0$. 
    (a) and (b) show implementations when either 1- or 2-qubit Pauli rotations are available in the set of hardware basis gates respectively.
    }
    \label{fig:ft_time_evol}
\end{figure}
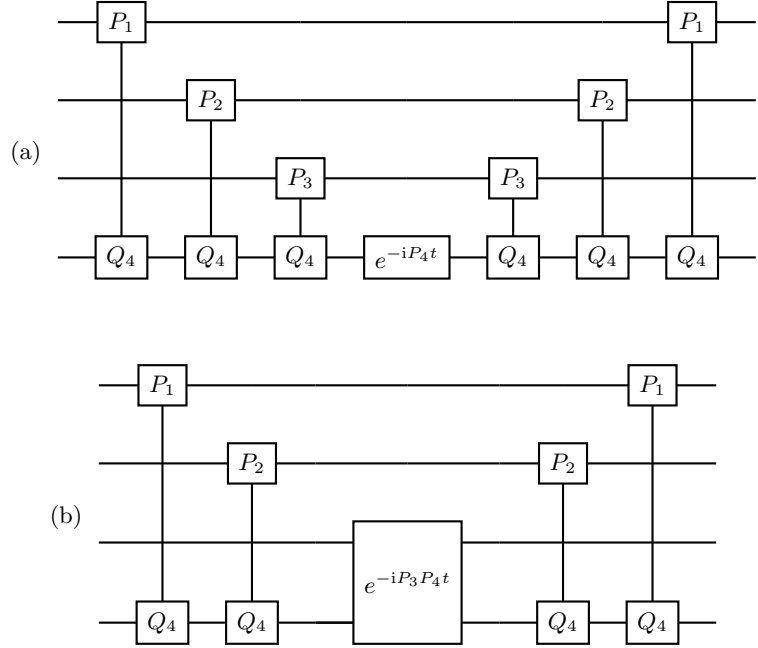

\begin{figure}
\centering
\begin{quantikz}
  \lstick{$q_{(0,0)}$} \qw & \qw & \qw & \qw & \qw & \qw & \qw & \qw \\
  \lstick{$q_{(0,1)}$} &\gate{Y} \vqw{7} & \qw &\qw & \qw &\qw &\qw & \qw  \\
  \lstick{$q_{(1,0)}$} & \qw & \gate{X} \vqw{6}& \qw & \qw & \qw & \qw & \qw  \\
  \lstick{$q_{(1,1)}$} & \qw & \qw & \gate{Z} \vqw{5}& \qw & \qw & \qw & \qw \\
 \lstick{$q_{(2,0)}$} & \qw & \qw & \qw & \gate{Y} \vqw{4} & \qw & \qw & \qw  \\
  \lstick{$q_{(2,1)}$} & \qw & \qw & \qw & \qw & \gate{X} \vqw{3}  & \qw & \qw \\
 \lstick{$q_{(3,0)}$} & \qw & \qw & \qw & \qw & \qw & \gate{Z} \vqw{2} & \qw  \\
 \lstick{$q_{(3,1)}$}\qw & \qw & \qw & \qw & \qw & \qw & \qw & \qw \\
 \lstick{$\ket{0}$} & \gate{X} & \gate{X} & \gate{X} & \gate{X} & \gate{X} & \gate{X} & \meter{} \\
\end{quantikz}
\caption{Syndrome measurement for the weight six loop operator $I_{(0,0)}Y_{(0,1)}X_{(1,0)}Z_{(1,1)}Y_{(2,0)}X_{(2,1)}Z_{(3,0)}I_{(3,1)}$.
Reproduced from Fig.~(5) in \cite{haggeErrorMitigationError2023}. 
}
\label{fig:stab_plaq}
\end{figure}
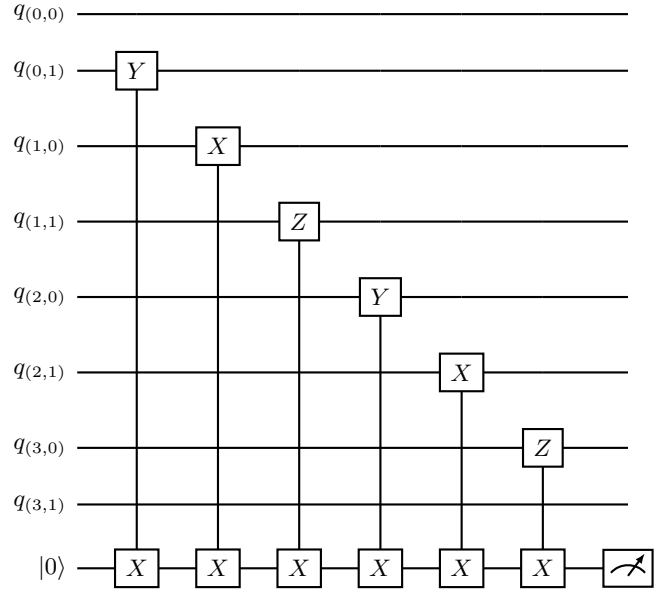

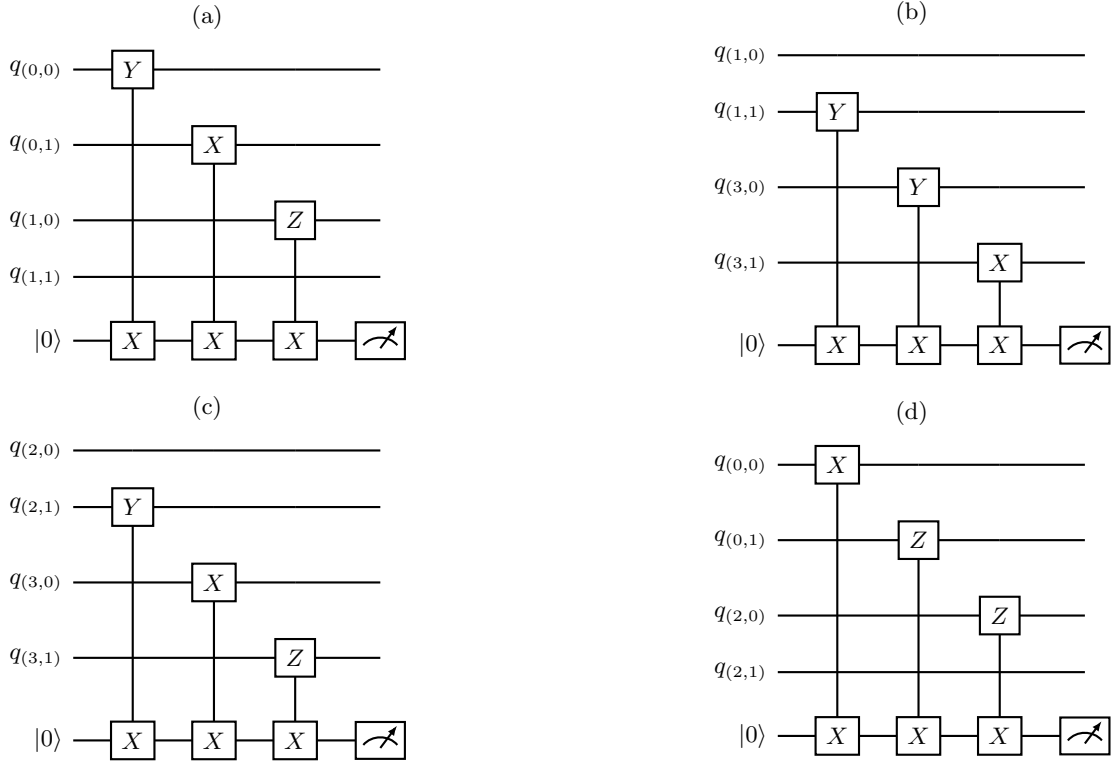
\begin{figure*}
\centering
\newcommand{\panel}[2]{\begin{minipage}{0.48\linewidth}\centering #1\\[4pt]#2\end{minipage}}
\panel{(a)}{\begin{quantikz}
  \lstick{$q_{(0,0)}$} & \gate{Y} \vqw{4} & \qw & \qw & \qw  \\
  \lstick{$q_{(0,1)}$} & \qw  & \gate{X} \vqw{3} &\qw & \qw \\
  \lstick{$q_{(1,0)}$} & \qw & \qw & \gate{Z} \vqw{2} & \qw  \\
  \lstick{$q_{(1,1)}$} & \qw & \qw & \qw & \qw \\
  \lstick{$\ket{0}$} & \gate{X} & \gate{X} & \gate{X}  & \meter{}
\end{quantikz}}\hfill
\panel{(b)}{\begin{quantikz}
  \lstick{$q_{(1,0)}$} & \qw & \qw & \qw & \qw  \\
  \lstick{$q_{(1,1)}$} &\gate{Y} \vqw{3} & \qw &\qw & \qw \\
  \lstick{$q_{(3,0)}$} & \qw & \gate{Y} \vqw{2}& \qw & \qw \\
  \lstick{$q_{(3,1)}$} & \qw & \qw & \gate{X} \vqw{1}& \qw \\
  \lstick{$\ket{0}$} & \gate{X} & \gate{X} & \gate{X}  & \meter{}
\end{quantikz}}\\[8pt]
\panel{(c)}{\begin{quantikz}
  \lstick{$q_{(2,0)}$} & \qw & \qw & \qw & \qw  \\
  \lstick{$q_{(2,1)}$} &\gate{Y} \vqw{3} & \qw &\qw & \qw \\
  \lstick{$q_{(3,0)}$} & \qw & \gate{X} \vqw{2}& \qw & \qw \\
  \lstick{$q_{(3,1)}$} & \qw & \qw & \gate{Z} \vqw{1}& \qw \\
  \lstick{$\ket{0}$} & \gate{X} & \gate{X} & \gate{X}   & \meter{}
\end{quantikz}}\hfill
\panel{(d)}{\begin{quantikz}
  \lstick{$q_{(0,0)}$} & \gate{X} \vqw{4} & \qw & \qw & \qw  \\
  \lstick{$q_{(0,1)}$} & \qw  & \gate{Z} \vqw{3} &\qw & \qw \\
  \lstick{$q_{(2,0)}$} & \qw & \qw & \gate{Z} \vqw{2} & \qw  \\
  \lstick{$q_{(2,1)}$} & \qw & \qw & \qw & \qw \\
  \lstick{$\ket{0}$} & \gate{X} & \gate{X} & \gate{X}  & \meter{}
\end{quantikz}}
\caption{Syndrome measurement for boundary loop operators (a) $-Y_{(0,0)}X_{(0,1)}Z_{(1,0)}I_{(1,1)}$, (b) $-I_{(1,0)}Y_{(1,1)}Y_{(3,0)}X_{(3,1)}$, (c) $-I_{(2,0)}Y_{(2,1)}X_{(3,0)}Z_{(3,1)}$, (d) $-X_{(0,0)}Z_{(0,1)}Z_{(2,0)}I_{(2,1)}$.}
\label{fig:stab_bound}
\end{figure*}

\begin{figure}
    \centering    
    \begin{subfigure}(a)
        \centering
        \begin{quantikz}
        & \gate{Z}\vqw{2} & \qw & \qw \\
        & \qw & \gate{Y}\vqw{1} & \qw \\
        \lstick{$\ket{0}$} & \gate{X} & \gate{X}  & \meter{}
        \end{quantikz}
    \end{subfigure}
    \\[1.5em]
    \begin{subfigure}(b)
        \centering
        \begin{quantikz}
        & \gate{Z}\vqw{4} & \qw & \qw & \qw & \qw & \qw & \qw \\
        & \qw & \gate{Y}\vqw{3} & \qw &  \qw  & \qw & \qw & \qw \\
        & \qw & \qw & \qw & \gate{Z}\vqw{2} & \qw & \qw &  \qw \\
        & \qw & \qw & \qw & \qw & \qw & \gate{Y}\vqw{1} & \qw  \\
        \lstick{$\ket{0}$} & \gate{X} & \gate{X} & \targ{} & \gate{X} & \targ{} & \gate{X}  & \meter{}\\
        \lstick{$\ket{+}$} & \qw & \qw & \ctrl{-1} & \qw  & \ctrl{-1}  & \gate{H} & \meter{}
        \end{quantikz}
    \end{subfigure}
    \caption{Quantum circuits for measuring weight-1 (a) and weight-2 (b) vertex operators. The weight-2 circuit includes a flag qubit to signal undetectable error propagation after a Z error on the ancilla after the second or 3rd Pauli controlled Pauli gates. (a) is reproduced from Fig.~(7) in \cite{haggeErrorMitigationError2023}.}
    \label{fig:vertex_meas}
\end{figure}
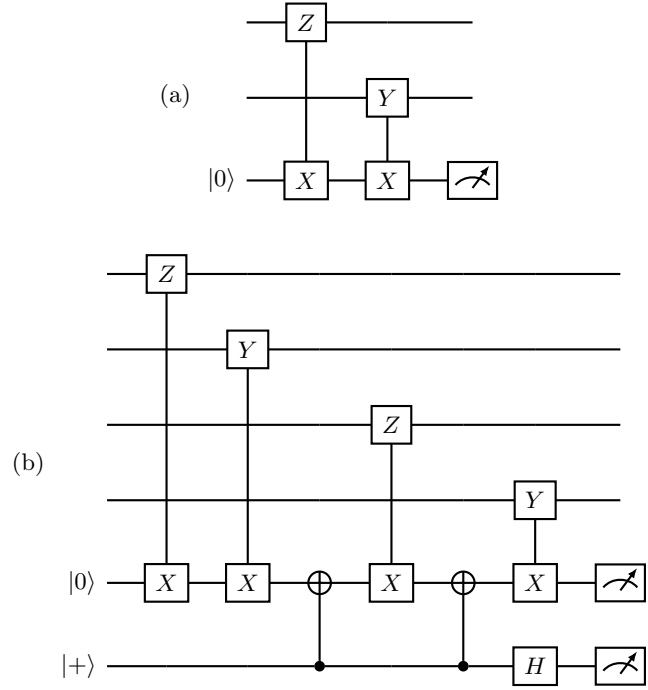

\section{Description of state preparation method} \label{app:prep}
In our simulations we prepare a vertex operator eigenstate based on the chosen fermion occupancy. We then measure each stabilizer once. If a stabilizer measurement $s_j$ returns -1, we apply a correction $R_j$ which anti-commutes with the stabilizer $S_j$ and commutes with $S_{i \neq j}$. This step ensures the state is a mutual +1 eigenstate of all the stabilizers. To then ensure the state has the correct vertex operator eigenvalues, for each vertex operator $B_k$, if the correction $R_j$ was applied and anti-commutes with $B_k$ we redefine the vertex operator $B_k \rightarrow -B_k$, otherwise $B_k$ remains unchanged. This redefinition is propagated through the circuit by classical feedforward of the initial stabilizer measurement outcomes. These are then used with the known $R_j$, $B_k$ commutation relations upon implementation of time evolution operators involving $B_k$. Measurements of $B_k$ are corrected classically using the initial stabilizer outcomes.

\section{\label{app:fh_circ} Circuits for simulating time evolution using generalized superfast encoding}

\subsection{Pauli-controlled-Pauli gates}
The circuits in the following sections make frequent use of Pauli-controlled-Pauli gates \cite{litinskiGameSurfaceCodes2019}. 
These are Clifford gates defined as
\begin{align}
    C(P_1, P_2) = e^{-\mathrm{i} (P_1 \otimes P_2 - I \otimes P_2  - P_1 \otimes I) \pi / 4}.
\end{align}
For example, a CNOT gate is $C(Z, X)$ up to a global phase.
The quantum circuit notation can be seen in Fig.~(\ref{fig:ft_time_evol})(a), where the first layer is a $C(P_1, Q_4)$ gate.

\subsection{Time evolution}
We implement approximate time evolution according to Eq.~(\ref{eq:trot}).
The basic circuit units of this decomposition are rotations of the form $U(P) = \exp(-\mathrm{i}P\theta)$, where $P \in \{B_jA_{(j,k)}, A_{(j,k)}B_k, B_j, B_jB_k \}$.
The Pauli weight of $P = P_1 P_2 P_3 P_4$ is between 2 and 4. 
In Fig.~(\ref{fig:ft_time_evol}) we show two circuits for implementing $U(P)$ using Pauli-controlled-Pauli gates, suited for different quantum hardware basis sets. If the weight of $P$ is less than 4, the same circuit is used with fewer Pauli-controlled-Pauli gates. 
Fig.~(\ref{fig:ft_time_evol})(a) gives a circuit for hardware which does not have 2-qubit Pauli rotations as native gates, in this case, $U(P)$ will not be fault-tolerant. 
However, the probability of an undetectable error being propagated can be reduced by implementing the following circuit modifications, which require classical propagation of single qubit errors through the circuits in Fig.~(\ref{fig:ft_time_evol}). 
\begin{enumerate}
    \item At each error location in $U(P)$, for each possible single qubit error $E$, choose $Q_4$ such that $[P_4, Q_4] \neq 0$ and, if possible, such that $E$ does not propagate through $U(P)$ to become undetectable. 
    \item $E$ may unavoidably propagate to become undetectable, or the choice of $Q_4$ may have been fixed by a previous error. If this is happens and $E$ does not act on the qubit targeted by the single-qubit rotation $\exp(-\mathrm{i} P_j t)$, then a flag qubit can be used to catch undetectable errors.
    \item If neither of the previous steps are possible, the operator $P$ should be evolved in the ``reflected'' code space. Reflection simply reverses the indices of $P$: $P_1 P_2 P_3 P_4 \rightarrow P_4 P_3 P_2 P_1$. Subsequent measurements of this operator should be consistent with this indexing.
\end{enumerate}
A more detailed explanation behind these modifications and why they eliminate instances of detectable errors propagating to undetectable errors can be found in Sec.~(5.1) of \cite{haggeErrorMitigationError2023}.
Following these steps will still leave cases where $E$ propagates to become $P$ acting after $U(P)$.
This case can be eliminated by using the circuit in Fig.~(\ref{fig:ft_time_evol})(b), which requires 2-qubit Pauli rotations to be native to the quantum hardware being used. 
Then each time evolution circuit $U(P)$ is fault-tolerant in the sense that detectable single qubit errors will not propagate to become undetectable.

\subsection{Stabilizer measurement}
Stabilizers can be measured without propagating detectable errors to undetectable errors using the circuits in Fig.~\ref{fig:stab_plaq} and Fig.~\ref{fig:stab_bound}, which are reproduced from~\cite{haggeErrorMitigationError2023}, where they show the fault-tolerant property by enumerating single qubit error propagations. Qubit indices are as described in Fig.~\ref{fig:index}

\subsection{Observable measurement}
\subsubsection{Fault-tolerant vertex operator measurement}
Measurement of lattice vertex operators is necessary to estimate Hamiltonian terms and important physical properties such as charge and spin correlators.
Here we show how to fault-tolerantly measure weight-one and -two vertex operators.

Weight one operators can be estimated using the circuit in Fig.~7 of \cite{haggeErrorMitigationError2023}, which we reproduce for convenience in Fig.~\ref{fig:vertex_meas}(a). 
This circuit is fault-tolerant in the sense that no single-qubit error arising before or during the circuit will propagate to an undetectable error. Any errors which do occur can be picked up by subsequent error detection circuits. 

Weight two operators measurement circuits are not immediately fault tolerant.
A Pauli $Z$ error acting on the ancilla qubit after the first or second Pauli controlled Pauli gates will propagate to a logical $ZY$ error on the second vertex. 
This type of error propagation can be detected by coupling the ancilla to a flag qubit~\cite{chaoQuantumErrorCorrection2018}.
With no noise the flag qubit will deterministically be measured in the $\ket{0}$ state and not affect the circuit output.
The Pauli $Z$ errors after the first or second layer which would cause undetectable errors will propagate 'down' the $CX$ gate and raise the flag. 
A $Z$ error after the first flag $CX$ will propagate to the 3rd vertex qubit, but will be caught by the second flag $CX$.
$X$ errors after either flag $CX$ or a $Z$ error after the second flag $CX$ will not propagate to undetectable errors but may trigger an incorrect flag event.

If this is the final measurement circuit involving the vertex subjected to a logical error, and vertex measurement will only be proceeded by error detection, there is no need for additional error handling, as the logical error will not affected data used for expectation value estimation, and will pass through subsequent detection circuits. 

\subsubsection{Trading flag qubits for additional samples}
In the previous section we demonstrated how to avoid the spread of undetectable errors by leveraging additional qubits and 2 qubit gates. 
We now outline how these additional qubits and gates can be replaced by using additional circuit samples and probabilistic error cancellation.

For the circuit in Fig.~\ref{fig:vertex_meas}(b) we could determine that 2 possible errors would propagate to be undetectable errors and to which errors they would propagate. 
To first order in physical error probability, the logical error channel for this circuit would be 
\begin{align}
   \Lambda (\rho) = (1 - p) \rho + p B_{j} \rho B_j,
\end{align}
where $j$ labels the second vertex and $p = p_{Z_{a, 1}} + p_{Z_{a, 2}}$ if $Z$ errors occur on the ancilla after the first and second layers with probability $p_{Z_{a, 1}}$ and $p_{Z_{a, 2}}$ respectively.
Therefore we can derive a subset of the logical noise model using more straightforwardly characterised physical error rates~\cite{bergProbabilisticErrorCancellation2022} and knowledge of Pauli error propagation through Clifford circuits. 
As these measurement circuits will contribute less to the circuit depth than circuits for time evolution, this logical error can be cancelled with a correspondingly small contribution to the total PEC overhead.

\end{document}